\documentclass[journal]{IEEEtran}

\usepackage{xcolor,soul,framed} 

\colorlet{shadecolor}{yellow}
\usepackage[pdftex]{graphicx}
\graphicspath{{../pdf/}{../jpeg/}}
\DeclareGraphicsExtensions{.pdf,.jpeg,.png}

\usepackage[cmex10]{amsmath}
\usepackage{array}
\usepackage{mdwmath}
\usepackage{mdwtab}
\usepackage{eqparbox}
\usepackage{url}

\newcolumntype{C}[1]{>{\centering\arraybackslash}p{#1}}
\usepackage{amssymb}
\usepackage{pifont}
\newcommand{\cmark}{\ding{51}}%
\newcommand{\xmark}{\ding{55}}%

\usepackage{longtable}
\usepackage{setspace}
\usepackage{hhline}
\usepackage{multicol}
\usepackage{multirow}

\begin{document}
\bstctlcite{IEEEexample:BSTcontrol}
    \title{Resilient Identification of Distribution Network Topology}
  \author{Mohammad~Jafarian,~\IEEEmembership{Member,~IEEE,}
      Alireza~Soroudi,~\IEEEmembership{Senior Member,~IEEE,}
      and Andrew~Keane,~\IEEEmembership{Senior Member,~IEEE}

  \thanks{ This work has emanated from research conducted with the financial support of Science Foundation Ireland under Grant No. SFI/16/IA/4496. The opinions, findings, and conclusions or recommendations expressed in this material are those of the author(s) and do not necessarily reflect the views of the Science Foundation Ireland.}
  \thanks{The authors are with the School of Electrical and Electronic Engineering, University College Dublin, Dublin 4, Ireland (e-mail: mohammad.jafarian@ucd.ie; alireza.soroudi@ucd.ie; andrew.keane@ucd.ie).}
 }

\markboth{
}{Roberg \MakeLowercase{\textit{et al.}}: High-Efficiency Diode and Transistor Rectifiers}

\maketitle

\begin{abstract}
Network topology identification (TI) is an essential function for distributed energy resources management systems (DERMS) to organize and operate widespread distributed energy resources (DERs). In this paper, discriminant analysis (DA) is deployed to develop a network TI function that relies only on the measurements available to DERMS. The propounded method is able to identify the network switching configuration, as well as the status of protective devices. Following, to improve the TI resiliency against the interruption of communication channels, a quadratic programming optimization approach is proposed to recover the missing signals. By deploying the propounded data recovery approach and Bayes' theorem together, a benchmark is developed afterward to identify anomalous measurements. This benchmark can make the TI function resilient against cyber-attacks. Having a low computational burden, this approach is fast-track and can be applied in real-time applications. Sensitivity analysis is performed to assess the contribution of different measurements and the impact of the system load type and loading level on the performance of the proposed approach.
\end{abstract}


%
\IEEEpeerreviewmaketitle

\section*{List of Abbreviations}
\addcontentsline{toc}{section}{List of Abbreviations}
\begin{IEEEdescription}[\IEEEusemathlabelsep\IEEEsetlabelwidth{AAAa}]
\item[DA]      Discriminant Analysis
\item[DERs]     Distributed Energy Resources
\item[DERMS]   DERs Management Systems
\item[DMS]     Distribution Management Systems
\item[ROC]     Receiver Operating Characteristic
\item[TI]      Topology Identification
\end{IEEEdescription}

\section{Introduction} \label{sec:introduction}
\IEEEPARstart{V}{arious} functions have been developed until now for the management of traditional distribution systems, with the main objective of distributing electricity among scattered loads. However, the increase in the penetration of distributed energy resources (DERs) has fundamentally altered the logic of these functions since their contribution to the supply of the demand can greatly influence the operation of distribution systems. Besides, the coordination of these units is not straightforward since each unit has its own unique static and dynamic characteristics \cite{Argonne}. This has led to a turning point, where, with the current design, distribution management systems (DMS) cannot exploit the capabilities of DERs in the management of distribution systems \cite{EPRI}.

\indent To address this issue, a variety of software-based solutions have been developed by the utilities and industry, referred to as distributed energy resources management systems (DERMS) \cite{8585806}, to cover the gap between managing the distribution system as a whole entity and managing the DERs \cite{EPRI-understanding}. The main objective of DERMS is to aggregate, monitor, and operate the widely dispersed DERs to provide maximum grid support \cite{DERMS_JOOS}.

\begin{figure}[tp]
\centerline{\includegraphics[width=\linewidth]{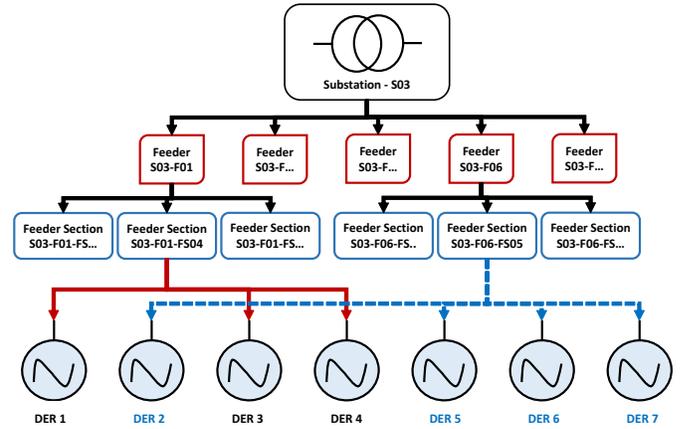}}
\caption{A typical distribution network organization hierarchy \cite{Argonne}.}
\label{fig:hierarchy}
\end{figure}

\indent Logically, access to the full-detail model of the distribution network is not necessary for DERMS, but it is crucial to have a simplified hierarchy of the network organization that addresses the feeder section, feeder circuit, and substation of the DERs \cite{Argonne,EPRI-understanding}. Fig. \ref{fig:hierarchy} presents a typical network organization hierarchy \cite{Argonne}. This organization hierarchy is not a static structure as it depends on the topology of the system, which is subjected to continuous changes due to the operation of protective devices or network reconfiguration operations for load transferring, planned outage services, load balancing, etc. \cite{Argonne}. Therefore, the identification of the real-time network topology is essential for DERMS to form the hierarchy organization and consequently, operate the DERs.

\indent Unlike transmission networks, the status of all the circuit breakers and protective devices is not available in distribution systems \cite{6175638}. Therefore, developing a function to process the network topology is crucial in distribution systems \cite{5424119}. Until now, different statistical and artificial intelligence techniques have been deployed for this purpose, as follows.
\begin{itemize}
  \item Maximum likelihood approach \cite{6175638}: this approach has been suggested for the identification of the network topology, by utilizing real-time measurements and the mean and covariance of the injected powers to all the network buses.
  \item Recursive Bayesian approach \cite{5424119}: the recursive Bayesian approach has been employed for the topology identification (TI) problem, by evaluating the state estimation error in all the possible network topologies, and predicting the most probable topology, in a recursive manner.
  \item Mixed-integer quadratic programming approach \cite{7027869}: a mixed-integer quadratic programming solution has been provided for the TI, based on minimizing the weighted square of the state estimation error. This approach is only applicable if the network is radially operated. In addition, the proposed formulation only applies to balanced networks.
  \item Nonconvex combinatorial optimization approach \cite{8754743}: the TI problem has been formulated as a nonconvex combinatorial optimization problem and a meter placement strategy has been adopted to ensure a unique solution. One of the drawbacks of this approach is that it does not support studying the weakly-meshed configurations. Furthermore, to employ this approach, meters should be of a specif type, i.e., measure the voltage and power transactions of system nodes together.
  \item Mixed-integer nonlinear programming approach \cite{8787584}: when only current measurements are available, it has been demonstrated that TI can be described as a mixed-integer nonlinear programming problem. This approach is based on minimizing the state estimation error. The application of this approach is restricted only to the networks that are equipped with line current sensors. If instead, other types of measurements are available, this approach will not provide a solution.
  \item Event-triggered approach \cite{7856295}: TI has been achieved by integrating an event-triggered TI stage to state estimation. This approach employs the recursive Bayesian approach of \cite{5424119} only to track the change of the network topology.
  \item Fuzzy logic model approach \cite{1490609}: a fuzzy logic model has been developed to identify the operation of protective devices. This approach does not provide the capability to identify the network switching configuration.
  \item Artificial neural network approach \cite{NN_topology_iden}: a feed-forward artificial neural network has been employed to evaluate the posterior probability of different network topologies, by applying the normalized exponential function as the output layer activation function.
\end{itemize}

\begin{table}[tp]
\centering
\caption{Capabilities and applications of different TI approaches.}
\scalebox{0.7}{\addtolength{\tabcolsep}{-3pt}
\begin{tabular}{C|c|c|c|c|c|c|c|c|c}
\hline
\multicolumn{1}{c||}{\textbf{Approach}$^{\mathrm{a}}$ $\rightarrow$} &
\multicolumn{1}{c|}{ML} & 
\multicolumn{1}{c|}{RB} &  
\multicolumn{1}{c|}{MIQP} &
\multicolumn{1}{c|}{NCO} &  
\multicolumn{1}{c|}{MINP} &
\multicolumn{1}{c|}{ET} & 
\multicolumn{1}{c|}{FL} & 
\multicolumn{1}{c|}{ANN} &
\multicolumn{1}{c}{DA} \\
\hhline{~~~~~~~~~~}
\multicolumn{1}{c||}{\textbf{Capability} $\downarrow$} &
\multicolumn{1}{c|}{\cite{6175638}} & 
\multicolumn{1}{c|}{\cite{5424119}} &  
\multicolumn{1}{c|}{\cite{7027869}} &
\multicolumn{1}{c|}{\cite{8754743}} &  
\multicolumn{1}{c|}{\cite{8787584}} &
\multicolumn{1}{c|}{\cite{7856295}} & 
\multicolumn{1}{c|}{\cite{1490609}} & 
\multicolumn{1}{c|}{\cite{NN_topology_iden}} &
\multicolumn{1}{c}{} \\
\hhline{----------}

\multicolumn{1}{C{3cm}||}{Supporting the weakly-meshed configuration} & 
\multicolumn{1}{C{0.7CM}|}{ \textcolor{white}{aaa} \cmark} &
\multicolumn{1}{C{0.7CM}|}{ \textcolor{white}{aaa} \cmark} &
\multicolumn{1}{C{0.7CM}|}{ \textcolor{white}{aaa} \xmark} &
\multicolumn{1}{C{0.7CM}|}{ \textcolor{white}{aaa} \xmark} &
\multicolumn{1}{C{0.7CM}|}{ \textcolor{white}{aaa} \cmark} &
\multicolumn{1}{C{0.7CM}|}{ \textcolor{white}{aaa} \cmark} &
\multicolumn{1}{C{0.7CM}|}{ \textcolor{white}{aaa} \cmark} &
\multicolumn{1}{C{0.7CM}|}{ \textcolor{white}{aaa} \cmark} &
\multicolumn{1}{C{0.7CM}}{ \textcolor{white}{aaa} \cmark} \\
\hline

\multicolumn{1}{C{3cm}||}{Applicable for unbalanced dist. networks} & 
\multicolumn{1}{C{0.7CM}|}{ \textcolor{white}{aaa} \cmark} &
\multicolumn{1}{C{0.7CM}|}{ \textcolor{white}{aaa} \cmark} &
\multicolumn{1}{C{0.7CM}|}{ \textcolor{white}{aaa} \xmark} &
\multicolumn{1}{C{0.7CM}|}{ \textcolor{white}{aaa} \cmark} &
\multicolumn{1}{C{0.7CM}|}{ \textcolor{white}{aaa} \cmark} &
\multicolumn{1}{C{0.7CM}|}{ \textcolor{white}{aaa} \cmark} &
\multicolumn{1}{C{0.7CM}|}{ \textcolor{white}{aaa} \cmark} &
\multicolumn{1}{C{0.7CM}|}{ \textcolor{white}{aaa} \cmark} &
\multicolumn{1}{C{0.7CM}}{ \textcolor{white}{aaa} \cmark} \\
\hline

\multicolumn{1}{C{3cm}||}{Applicable for different types of measurements} & 
\multicolumn{1}{C{0.7CM}|}{ \textcolor{white}{aaa} \cmark} &
\multicolumn{1}{C{0.7CM}|}{ \textcolor{white}{aaa} \cmark} &
\multicolumn{1}{C{0.7CM}|}{ \textcolor{white}{aaa} \cmark} &
\multicolumn{1}{C{0.7CM}|}{ \textcolor{white}{aaa} \xmark} &
\multicolumn{1}{C{0.7CM}|}{ \textcolor{white}{aaa} \xmark} &
\multicolumn{1}{C{0.7CM}|}{ \textcolor{white}{aaa} \cmark} &
\multicolumn{1}{C{0.7CM}|}{ \textcolor{white}{aaa} \cmark} &
\multicolumn{1}{C{0.7CM}|}{ \textcolor{white}{aaa} \cmark} &
\multicolumn{1}{C{0.7CM}}{ \textcolor{white}{aaa} \cmark} \\
\hline

\multicolumn{1}{C{3cm}||}{Identifying the switching configurations} & 
\multicolumn{1}{C{0.7CM}|}{ \textcolor{white}{aaa} \cmark} &
\multicolumn{1}{C{0.7CM}|}{ \textcolor{white}{aaa} \cmark} &
\multicolumn{1}{C{0.7CM}|}{ \textcolor{white}{aaa} \cmark} &
\multicolumn{1}{C{0.7CM}|}{ \textcolor{white}{aaa} \cmark} &
\multicolumn{1}{C{0.7CM}|}{ \textcolor{white}{aaa} \cmark} &
\multicolumn{1}{C{0.7CM}|}{ \textcolor{white}{aaa} \cmark} &
\multicolumn{1}{C{0.7CM}|}{ \textcolor{white}{aaa} \xmark} &
\multicolumn{1}{C{0.7CM}|}{ \textcolor{white}{aaa} \cmark} &
\multicolumn{1}{C{0.7CM}}{ \textcolor{white}{aaa} \cmark} \\
\hline

\multicolumn{1}{C{3cm}||}{Identifying the status of protective devices} & 
\multicolumn{1}{C{0.7CM}|}{ \textcolor{white}{aaa} \cmark} &
\multicolumn{1}{C{0.7CM}|}{ \textcolor{white}{aaa} \cmark} &
\multicolumn{1}{C{0.7CM}|}{ \textcolor{white}{aaa} \cmark} &
\multicolumn{1}{C{0.7CM}|}{ \textcolor{white}{aaa} \cmark} &
\multicolumn{1}{C{0.7CM}|}{ \textcolor{white}{aaa} \cmark} &
\multicolumn{1}{C{0.7CM}|}{ \textcolor{white}{aaa} \cmark} &
\multicolumn{1}{C{0.7CM}|}{ \textcolor{white}{aaa} \cmark} &
\multicolumn{1}{C{0.7CM}|}{ \textcolor{white}{aaa} \cmark} &
\multicolumn{1}{C{0.7CM}}{ \textcolor{white}{aaa} \cmark} \\
\hline

\multicolumn{10}{l}{$^{\mathrm{a}}$ML = Maximum Likelihood, RB = Recursive Bayesian,} \\
\hhline{~~~~~~~~~~}
\multicolumn{10}{l}{MIQP = Mixed-Integer Quadratic Programming,} \\
\hhline{~~~~~~~~~~}
\multicolumn{10}{l}{NCO = Nonconvex Combinatorial Optimization,} \\
\hhline{~~~~~~~~~~}
\multicolumn{10}{l}{MINP = Mixed-Integer Nonlinear Programming,} \\
\hhline{~~~~~~~~~~}
\multicolumn{10}{l}{ET = Event Triggered, FL = Fuzzy Logic, ANN = Artificial Neural Network.} \\
\hhline{~~~~~~~~~~}
\end{tabular}} \label{tab:comparision_capabilities}
\end{table}

\indent Table \ref{tab:comparision_capabilities} summarizes the capabilities and applications of these approaches. As noted, the applications of some of them are restricted to particular distribution systems. Besides, most of these approaches are based on minimizing the state estimation error, which requires accessing the full network model and pseudo-measurements.
Since these approaches are originally developed for DMS, having access to this information has been presumed. In the case of DERMS, however, the availability of this information is debatable \cite{Argonne,EPRI-understanding}.
It should also be remarked that in distribution systems, the quality of demand estimation is poor and pseudo-measurements generally contain a high level of uncertainty \cite{4810103}. This situation has been aggravated by the wide utilization of electric vehicles and flexible demands, which introduce further uncertainty to the system. Therefore, even if available, pseudo-measurements are not reliable and it is beneficial if TI can be implemented without relying on them.

\indent While deploying information technologies streamlines the system management, it can cause more dependencies between the physical and cyber components, which makes the system more vulnerable to natural disasters and cyber-attacks \cite{ARGHANDEH20161060}. In fact, proneness to loss of online data has undermined the popularity of the approaches that rely on online measurements in practice, especially when the system may face the loss of multiple signals, because of, e.g., interruption of communication channels, or cyber-attacks.
Securing the power system management against cyber-attacks has been the subject of many recent studies, like \cite{hooshyar1} and \cite{hooshyar2}.
This highlights the importance of resiliency in the development of a TI function. None of the reviewed TI approaches is resilience-oriented.

\indent This paper is dedicated to developing a topology identification function for distribution networks that relies only on the measurements available to DERMS.
These measurements include the operating condition of the grid and DERs \cite{Argonne}.
Considering the applicability of discriminant analysis (DA) as a widely-used statistical classifier, DA is employed for this purpose.
While originally developed for DERMS, this method is applicable for DMS too since DMS and DERMS share the DERs measurements \cite{Argonne}.
To enhance the resiliency of the proposed TI function against the interruption of communication channels, a quadratic programming optimization approach is proposed to recover the missing signals.
Following, by deploying this data recovery approach and Bayes' theorem, a benchmark is introduced to detect if some measurements contain anomalous values. This benchmark makes the proposed TI function resilient against cyber-attacks.
This approach requires a low processing time, which makes it suitable to be employed in real-time applications. 
This approach is able to consider weakly-meshed configurations, works with any type of measurements, can treat unbalanced networks, and can be applied to identify the switching configuration and the operation of protective devices.

\indent Following, the application of DA to TI is elaborated in Section \ref{sec:Methodology}. In this section also an optimization-based approach is proposed to enhance the resiliency of the TI function. The propounded method is examined on a modified version of the IEEE 123 node test feeder. This test system is described in Section \ref{sec:study System} and the results are discussed in Section \ref{sec:simulation Study}. Sensitivity analysis is conducted in Section \ref{sec:sensitivity}. Section \ref{sec:conclusion} concludes the paper.

\section{Methodology} \label{sec:Methodology}
\indent Since network topologies are distinctive, TI is basically a classification problem. Classification techniques are categorized into statistical and artificial intelligence techniques. 
DA is a widely used statistical classifier \cite{most_common} that has already proven its capability in different fields of science, e.g., financial studies \cite{SERRANOCINCA20131245}, biomedical studies \cite{6508865}, earth science \cite{CHIEN2018379}, etc.
DA has also been applied in the field of power system studies, from load forecasting \cite{49084}, load modeling \cite{6476045}, and fault detection \cite{7066231}, to assessing the power system security \cite{153398,962423}.


\indent Let $X$ be an observation that contains $n$ predictors as $X=[x_1,...,x_n]$ (a predictor is a system's feature that is employed for the classification and an observation is a set that contains all the predictors). Suppose $X$ may belong to $K$ different classes and $\rho_k$ is the prior probability of $X$ associating with the $k$th class (the portion of the observations in the training set that are associated with the $k$th class). Assume that $f_k(X)$ is the density function of the $k$th class. The Bayes' theorem states the probability of $X$ associating with the $k$th class, $p(class=k|X)$, as \cite{James}:
\begin{equation}
\label{eq:bayes}
p(class=k|X)=(f_k(X)\rho_k)/\displaystyle\sum^K_{i=1}f_i(X)\rho_i
\end{equation}

\indent DA assumes a multivariate normal distribution for each class as:
\begin{equation}
\label{eq:gaussian}
f_k(X)=\frac{exp \big( -0.5(X-M_k)\Sigma^{-1}_k(X-M_k)^T \big) }{(2\pi|\Sigma_k|)^{0.5}}
\end{equation}
where $M_k$ ($1 \times n$) and $\Sigma_k$ ($n \times n$) are the mean and covariance of the normal distribution for the $k$th class, respectively. \indent Classification to the largest probability will make the smallest expected number of misclassifications \cite{James}. Therefore, DA assigns each observation to the class with the largest probability. This is equivalent to assigning to the class with the largest $\delta_k(X)$, as \cite{james2013introduction}:

\begin{equation}
\label{eq:delta}
\begin{split}
\delta_k(X)=&-0.5(X-M_k)\Sigma_k^{-1}(X-M_k)^T\\
&-0.5log|\Sigma_k|+log\rho_k.
\end{split}
\end{equation}

In this paper, maximum a posteriori estimation \cite{38136} is employed to form the Discriminant analysis model in the training phase, i.e., determining the mean vector and the covariance matrix concerning each network topology. While the normal distribution is assumed in DA, it has been demonstrated that DA is robust to the violation of this assumption \cite{robust_1}, especially when predictors are continuous and bounded \cite{robust_2}, like in the case of distribution systems. 


\subsection{Discriminant Analysis for Topology Identification} \label{sec:proposed method}
The topology of a distribution system changes mainly due to network reconfiguration operations and faults. Therefore, the proposed method is supposed to identify the network switching configuration and the status of protective devices, employing the measurements available to DERMS. In this regard, it is supposed that DERs and the MV-side of the HV/MV substation are equipped with PQV sensors (that measure the active power ($P$), reactive power ($Q$), and the connecting voltage, in the case of DERs, and the MV-side voltage, in the case of the substation ($V$)) \cite{6204095}. The distribution system may be unbalanced and contain both the three- and one-phase parts. It may also contain auto-regulators, which make the identification process more arduous since they apply control actions on the system that DERMS might be ignorant of.

\indent Without loss of generality, it is assumed that DERs are operated under a constant power factor (Irish regulations \cite{IrelandGridCode}). Therefore, the DERs reactive power generation does not contain new information; also since the network is unbalanced, the negative sequence component of the voltage should be considered in addition to the positive sequence component, resulting in three available measurements from each of the DERs: $P_{DER-j}$ (the active power generation of the $j$th DER), $V_{DER-j}^+$, and  $V_{DER-j}^-$ (the positive and negative sequence components of the $j$th DER voltage, respectively). From the substation, four measurements are available: $P_{sub}$, $Q_{sub}$ (the active and reactive powers, respectively), $V_{sub}^+$, and $V_{sub}^-$ (the positive and negative sequence components of the MV-side voltage, respectively).

It should be emphasized that in this approach, we do not consider the objective of network reconfiguration operations, but instead, to form the training and test data sets, all the possible configurations are considered and in each configuration, all the probable variations of loads consumption and DERs generation are simulated. In this order, whatever the objective of network configuration is, it will include a subset of this data set.

\subsection{Enhancing The Resiliency}
As discussed, the loss of multiple signals and malicious data are two phenomena that jeopardize the applicability of approaches relying on online measurements in practice. In this section, an approach is proposed to recover the original values of the measurements that are missing or suspected to contain anomalous values. From now on, they are referred to as the missing signals.

\indent Suppose $X=[...,x^u_1,...,x^u_2,...,x^u_l,...]$ is an observation that belongs to the $k$th class (that we are unaware of) and it contains $n$ predictors. Let $X^u=[x^u_1,...,x^u_l]$ contain the missing signals and  $X^v=[x^v_1,...,x^v_q]$ include the available signals, with $l$ the number of missing signals and q the number of available signals, such that $l+q=n$. Assume that the minimum ($X^u_{min}=[min(x^u_1),...,min(x^u_l)]$) and maximum ($X^u_{max}=[max(x^u_1),...,max(x^u_l)]$) of the missing signals are known (considering the range of their variations in the training set). In the training process, DA looks for $M$ and $\Sigma$ which maximize $\delta$ in \eqref{eq:delta} for the correct class; here, $M$ and $\Sigma$ are given, but some predictors are missing. The idea is to determine the values of the missing signals such that $\delta_k$ gets to its maximum for the $k$th class, subject to the missing signals being confined to their known range of variations as:
\begin{equation}
\label{eq:max}
\begin{split}
\max_{X^{u}}  \hspace{2mm}& \delta_k([..., x^u_1,...,x^u_l, ... ]  \\
 Subject& \mbox{  }to: \mbox{  } X^{u}_{min} \leq X^{u} \leq X^{u}_{max}
\end{split}
\end{equation}

Substituting $\delta_k(x)$ from \eqref{eq:delta} into \eqref{eq:max} gives:
\begin{equation}
\label{eq:substitute}
\begin{split}
\max_{X^{u}} \hspace{2mm}& -0.5([...,x^u_1,...,x^u_l,...]-M_k)\Sigma_k^{-1} \\
&([...,x^u_1,...,x^u_l,...]-M_k)^T\\
&-0.5log|\Sigma_k|+log\rho_k.\\
 Subject& \mbox{  }to: \mbox{  } X^{u}_{min} \leq X^{u} \leq X^{u}_{max}
\end{split}
\end{equation}

As $|\Sigma_k|$ and $\rho_k$ are independent of $X^u$, they can be omitted in \eqref{eq:substitute}. In addition, considering the minus sign, \eqref{eq:substitute} can be rewritten as:
\begin{equation}
\label{eq:substitute_simple}
\begin{split}
\min_{X^{u}} \hspace{2mm}& 0.5([...,x^u_1,...,x^u_l,...]-M_k)\Sigma_k^{-1} \\
&([...,x^u_1,...,x^u_l,...]-M_k)^T\\
 Subject& \mbox{  }to: \mbox{  } X^{u}_{min} \leq X^{u} \leq X^{u}_{max}
\end{split}
\end{equation}

Suppose $X$ is rephrased to form $X^{reph}=[X^u,X^v]$, such that the available and missing signals are separated. Suppose also $M_k$ and $\Sigma_k$ are rephrased in compliance to form $M^{reph}_k=[M^u_k,M^v_k]$ and $\Sigma^{reph}_k$, respectively. Let $\Psi_k$ denote the inverse of $\Sigma^{reph}_k$ as:
\begin{equation}
\label{eq:sigma_inverse}
(\Sigma^{reph}_k)^{-1}=\Psi_k =
\begin{bmatrix}
  \Psi^{uu}_k & \Psi^{uv}_k\\
  \Psi^{vu}_k & \Psi^{vv}_k
\end{bmatrix}
\end{equation}
where $\Psi^{uu}_k$, $\Psi^{uv}_k$, $\Psi^{vu}_k$, and $\Psi^{vv}_k$ are $l \times l$, $l \times q$, $q \times l$, and $q \times q$ matrices, respectively.

\indent Considering \eqref{eq:sigma_inverse} and the rephrased terms, \eqref{eq:substitute_simple} can be presented as:
\begin{equation}
\label{eq:substitute_rephrased}
\begin{split}
\min_{X^{u}} \hspace{2mm}& 0.5([X^u-M^u_k]\Psi^{uu}_k[X^u-M^u_k]^T\\
&+[X^u-M^u_k]\Psi^{uv}_k[X^v-M^v_k]^T \\
&+[X^v-M^v_k]\Psi^{vu}_k[X^u-M^u_k]^T\\
&+[X^v-M^v_k]\Psi^{vv}_k[X^v-M^v_k]^T)\\
 Subject& \mbox{  }to: \mbox{  } X^{u}_{min} \leq X^{u} \leq X^{u}_{max}
\end{split}
\end{equation}

 Since $X^u\Psi^{uv}_k[X^v-M^v_k]^T$ and $X^u\Psi^{uu}_k(M^u_k)^T$ are of scalar type, we have:
\begin{equation}
\label{eq:transpose}
\begin{split}
X^u\Psi^{uv}_k[X^v-M^v_k]^T &= [X^v-M^v_k](\Psi^{uv}_k)^T(X^u)^T \\
X^u\Psi^{uu}_k(M^u_k)^T &= M^u_k(\Psi^{uu}_k)^T(X^u)^T
\end{split}
\end{equation}

\indent By considering \eqref{eq:transpose} and omitting the parts that are independent of $X^u$, \eqref{eq:substitute_rephrased} is simplified as:
\begin{equation}
\label{eq:quadprog}
\begin{split}
\min_{X^{u}}  \hspace{2mm}& 0.5X^{u}\Psi^{uu}_k(X^{u})^T+R_k(X^{u})^T \\
 Subject& \mbox{  }to: \mbox{  } X^{u}_{min} \leq X^{u} \leq X^{u}_{max}
\end{split}
\end{equation}
where $R_k$ is defined as:
\begin{equation}
\label{R} 
\begin{split}
R_k &= 0.5\big( [X^v-M^v_k](\Psi_k^{vu}+(\Psi^{uv}_k)^T)-\\
&M^u_k(\Psi^{uu}_k+(\Psi^{uu}_k)^T) \big)
\end{split}
\end{equation}
\indent Equation \eqref{eq:quadprog} is an $l$-variable quadratic programming problem that can be solved by setting the gradient of the objective function equal to zero and checking the positive definiteness of the Hessian matrix ($\Psi^{uu}_k$) and lower/upper bounds. Since the true class of $X$ is unknown, \eqref{eq:quadprog} should be solved for all the classes, resulting in $K$ sets of values for the missing signals. Now, the maximum amount of \eqref{eq:delta} with different sets represents the most probable class and hence, the best matching set of recovered signals.

\subsection{Anomaly Detection}

In addition to the loss of multiple signals, the TI approach should be resilient against malicious data. In this subsection, a benchmark is introduced to detect if one or a group of measurements contain anomalous values. Suppose the measurements from a specific meter, i.e., $X^s=[x_1^s,...x_l^s]$, are suspected to contain anomalous values. Considering \eqref{eq:quadprog}, the information of the other measurements can be exploited to estimate the normal value of $X^s$, i.e., $X^r=[x_{1}^r,...x_{l}^r]$. To discern if $X^s$ is anomalous, the idea is to compare the likelihood of $X^r$ associating with any of the network topologies, i.e., $\cup^K_{i=1}(k_i)$, to that of $X^s$, where $\cup^K_{i=1}(k_i)$ denotes the union of all the possible topologies. For this purpose, the likelihood ratio, denoted by $\alpha=\Lambda(X^r:X^s|\cup^K_{i=1}(k_i))$, is employed as a benchmark to detect if $X^s$ is anomalous. The likelihood ratio is defined as:
\begin{equation} \label{eq:likelihood_ratio}
\alpha=\Lambda(X^r:X^s|\cup^K_{i=1}(k_i))=
\frac{p(\cup^K_{i=1}(k_i)|X^r)}{p(\cup^K_{i=1}(k_i)|X^s)}
\end{equation}

\indent If $X^s$ is anomalous, the probability of $X^s$ to be associated with any of the topologies would be small, which results in high values for $\alpha$. If, on the other hand, $X^s$ is normal, $\alpha$ is expected to take smaller values (close to 1).

Since the network topologies are distinctive, \eqref{eq:likelihood_ratio} can be rewritten as:
\begin{equation} \label{eq:likelihood_ratio2}
\alpha=
\big( \displaystyle\sum_{i=1}^K p(k_i|X^r) \big) /
\big( \displaystyle\sum_{i=1}^K p(k_i|X^s) \big)
\end{equation}

\indent  According to Bayes' theorem, $p(k_i|X)$ is equal to $p(X|k_i)p(k_i)/p(X)$. Therefore, we have:
\begin{equation} \label{eq:likelihood_ratio3}
\alpha=
\big( \displaystyle\sum_{i=1}^K f_k(X^r)\rho_k/p(X^r) \big) /
\big( \displaystyle\sum_{i=1}^K f_k(X^s)\rho_k/p(X^s) \big)
\end{equation}

\indent Assuming that there is no privilege to choose $X^r$ over $X^s$, or in other words, no prior judgment is made about $X^s$ being anomalous, $p(X^s)/p(X^r)$ would be equal to one, which gives the following benchmark to detect if $X^s$ is anomalous.
\begin{equation} \label{eq:likelihood_ratio4}
\alpha=
\big( \displaystyle\sum_{i=1}^K f_k(X^r)\rho_k \big) /
\big( \displaystyle\sum_{i=1}^K f_k(X^s)\rho_k \big)
\end{equation}

It is worth emphasizing that the quadratic programming data recovery approach is applicable only when some measurements are not available or contain anomalous values. In this case, first, \eqref{eq:quadprog} is used to predict the original values of the missing or anomalous measurements and then \eqref{eq:delta} is applied to predict the network topology with the recovered measurements values. Should it be not the case, \eqref{eq:delta} is directly applied using the original measurements values.

\begin{figure}[tp]
\centerline{\includegraphics[width=\linewidth]{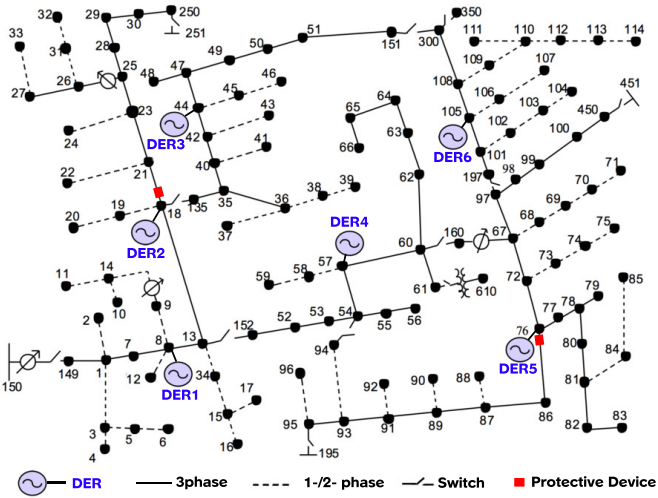}}
\caption{The IEEE 123 node test feeder, integrated with 6 DERs.}
\label{fig:fig_IEEE123}
\end{figure}

\begin{table}[tp]
\caption{Network topologies.}
\scalebox{0.8}{\addtolength{\tabcolsep}{-1pt}
\begin{tabular}{c|c|c|c|c|c|c}
\hline
\multicolumn{1}{c||}{} & 
\multicolumn{6}{c}{  \textbf{Switching configurations}} \\
\cline{2-7} 
\multicolumn{1}{c||}{  \textbf{Switches}} & 
\multicolumn{1}{c|}{  C1} & 
\multicolumn{1}{c|}{  C2} & 
\multicolumn{1}{c|}{  C3} & 
\multicolumn{1}{c|}{  C4} & 
\multicolumn{1}{c|}{ \hspace{1mm} C5 \hspace{1mm}} & 
\multicolumn{1}{c}{ \hspace{1mm} C6 \hspace{1mm}} \\
\hhline{-------}
\multicolumn{1}{c||}{ $S_{13-152}$} & 
\multicolumn{1}{c|}{  0$^{\mathrm{a}}$} & 
\multicolumn{1}{c|}{  0} & 
\multicolumn{1}{c|}{  \textbf{\textcolor[HTML]{FF0000}{1}}} & 
\multicolumn{1}{c|}{  0} & 
\multicolumn{1}{c|}{  0} & 
\multicolumn{1}{c}{  0} \\
\hhline{~~~~~~~}
\multicolumn{1}{c||}{  $S_{60-160}$} & 
\multicolumn{1}{c|}{  0} & 
\multicolumn{1}{c|}{  0} & 
\multicolumn{1}{c|}{  0} & 
\multicolumn{1}{c|}{  \textbf{\textcolor[HTML]{FF0000}{1}}} & 
\multicolumn{1}{c|}{  0} & 
\multicolumn{1}{c}{  0} \\
\hhline{~~~~~~~}
\multicolumn{1}{c||}{  $S_{97-197}$} & 
\multicolumn{1}{c|}{  0} & 
\multicolumn{1}{c|}{  0} & 
\multicolumn{1}{c|}{  0} & 
\multicolumn{1}{c|}{  0} & 
\multicolumn{1}{c|}{  \textbf{\textcolor[HTML]{FF0000}{1}}} & 
\multicolumn{1}{c}{  0} \\
\hhline{~~~~~~~}
\multicolumn{1}{c||}{  $S_{151-300}$} & 
\multicolumn{1}{c|}{  \textbf{\textcolor[HTML]{FF0000}{1}}} & 
\multicolumn{1}{c|}{  0} & 
\multicolumn{1}{c|}{  0} & 
\multicolumn{1}{c|}{  0} & 
\multicolumn{1}{c|}{  0} & 
\multicolumn{1}{c}{  0} \\
\hhline{~~~~~~~}
\multicolumn{1}{c||}{  $S_{18-135}$} & 
\multicolumn{1}{c|}{  0} & 
\multicolumn{1}{c|}{  0} & 
\multicolumn{1}{c|}{  0} & 
\multicolumn{1}{c|}{  0} & 
\multicolumn{1}{c|}{  0} & 
\multicolumn{1}{c}{  \textbf{\textcolor[HTML]{FF0000}{1}}} \\
\hhline{-------}
\multicolumn{1}{c||}{} & 
\multicolumn{4}{c}{  \textbf{Status of protective devices}} & \multicolumn{1}{c}{ } &\\
\hhline{~----~~}

\multicolumn{1}{c||}{  \textbf{P. devices}} & 
\multicolumn{1}{c|}{  Cx-00} & 
\multicolumn{1}{c|}{  Cx-01} & 
\multicolumn{1}{c|}{  Cx-10} & 
\multicolumn{1}{c}{  Cx-11} & \multicolumn{1}{c}{ } &  \\
\hhline{-----~~}
\multicolumn{1}{c||}{  $PD_{18}$} & 
\multicolumn{1}{c|}{  0} & 
\multicolumn{1}{c|}{  0} & 
\multicolumn{1}{c|}{  \textbf{\textcolor[HTML]{FF0000}{1}}} & 
\multicolumn{1}{c}{  \textbf{\textcolor[HTML]{FF0000}{1}}} &
\multicolumn{1}{c}{ } 
& \\
\hhline{~~~~~~~}
\multicolumn{1}{c||}{ $PD_{76}$} & 
\multicolumn{1}{c|}{  0} & 
\multicolumn{1}{c|}{  \textbf{\textcolor[HTML]{FF0000}{1}}} & 
\multicolumn{1}{c|}{  0} & 
\multicolumn{1}{c}{  \textbf{\textcolor[HTML]{FF0000}{1}}} &
\multicolumn{1}{c}{ } & \\
\hhline{-------}
\multicolumn{7}{l}{ $^{\mathrm{a}}$0 represents close; \textbf{\textcolor[HTML]{FF0000}{1}} represents open.}
\end{tabular}}
 			\label{tab:scenarios}
 \end{table}

\section{Study System} \label{sec:study System}

The IEEE 123 node test feeder in Fig. \ref{fig:fig_IEEE123} is used as the test system \cite{119237}. This system operates at a nominal voltage of 4.16 kV and includes 3-phase, 2-phase, and 1-phase overhead lines and underground cables. This system also contains four shunt capacitor banks and four auto-regulators. Loads are unbalanced with constant current, impedance, and power. There are 16 switches in this system, five of which are used for network reconfiguration operations: $S_{13-152}$, $S_{60-160}$, $S_{97-197}$, $S_{151-300}$, and $S_{18-135}$ ($S_{94-54}$ is a one-phase switch and as its operation insignificantly changes the system operating condition, it is not considered here).

\indent To simulate a high DERs penetrated system, the test system is modified by placing six DERs at buses 8, 18, 44, 57, 76, and 105 (DER1 to DER6 respectively) with the mean active power generation of 100, 100, 70, 200, 300, and 80 kW, respectively (about 7\% penetration). Defining a reconfiguration zone as an artificial area surrounded by the network switches, DERs are located such that in each zone at least one of the DERs (and hence one measurement point) exists. It will be discussed later that the proposed approach is still potent, in case this assumption is neglected; also since DERMS are purposeful for a distribution system with a noticeable DERs penetration, this assumption is practical. Two protective devices are also placed at buses 18 and 76 ($PD_{18}$ and $PD_{76}$).

\begin{figure}[tp]
\centering
\includegraphics[width=\linewidth]{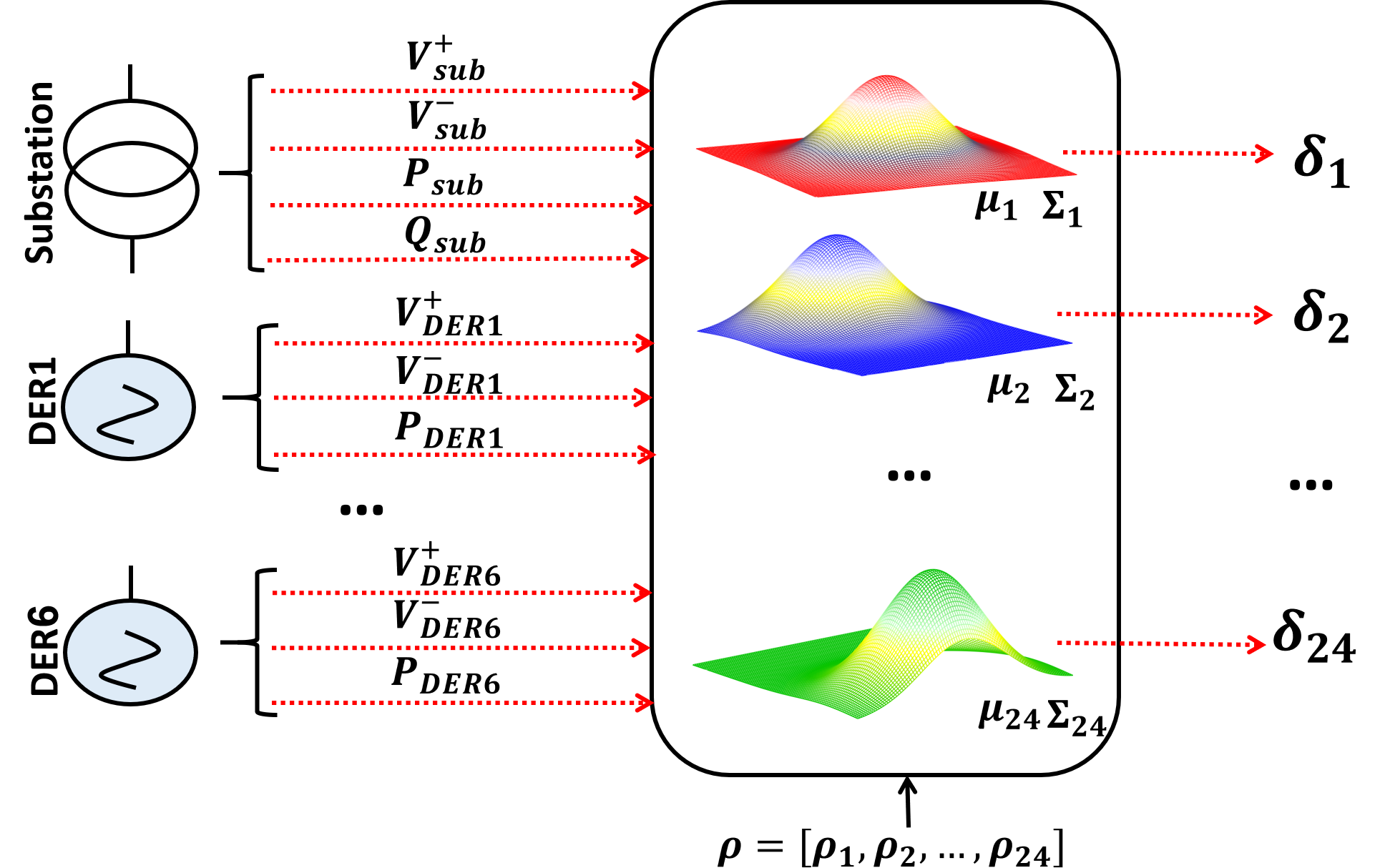}
\caption{A schematic diagram of the proposed DA approach.}
\label{fig:schematic}
\end{figure}

\indent Considering the network switches, six possible configurations do not lead to the interruption of the system loads. They are labeled as configuration 1 (C1) to configuration 6 (C6). Table \ref{tab:scenarios} presents the details. Only in C2 the network is weakly-meshed, while in the other five configurations, the network is radial. In each switching configuration, four statuses can be assumed for the two protective devices, resulting in a total of 24 possible topologies, and hence, classes to be identified. Fig. \ref{fig:schematic} depicts a schematic diagram for the proposed DA approach.

\begin{figure}[tp]
\centering
\includegraphics[width=\linewidth]{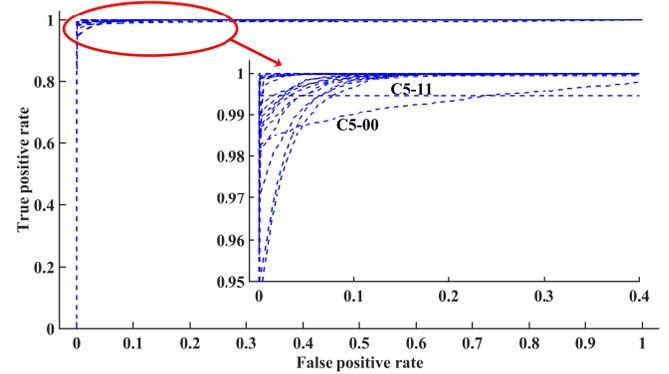}
\caption{ROC curves for the proposed DA approach with the training set.}
\label{fig:ROC}
\end{figure}

\section{Simulation Results} \label{sec:simulation Study}
In this section, first, the results of DA and then the results of the proposed data recovery approach are presented.
\subsection{Topology Identification}
Monte Carlo simulations (mean=1 and standard deviation=0.3) are used to explore different load values (15000 simulations per topology and 360000 simulations cumulatively). In these simulations, to capture the variations in the DERs active power generation, in addition to the loads active power consumption, the active power generation of DERs is varied too, within their range of variation. The high number of simulations ensures that all the variations of the DERs active power generation are considered. As mentioned before, due to the Irish regulations, the DERs power factor is assumed unchanged (0.95 lead). Another option is to consider different set points for the DERs, by sweeping their active power generation from zero to the rated power, and for each set point, Monte Carlo simulations are conducted. As we tried this scenario, no noticeable change was observed in the performance of the proposed approach.
The measurement noise is simulated as an independent zero-mean Gaussian with a standard deviation of 1\%. 90\% of the simulations are used to train DA and the remainder 10\% are used to form the test set (per configuration).

\indent Receiver operating characteristic (ROC) analysis is a common tool to evaluate the performance of classifiers in the training phase \cite{5419278}. In this analysis, in spite of 0.5, other thresholds (a real parameter between zero and one) are considered for the classification in \eqref{eq:bayes} and the true-positive rate (fraction of correct classification) versus false-positive rate (fraction of misclassification) are depicted at different thresholds for each class. The area under a ROC curve, $AUC$, indicates the overall performance of a classifier. An ideal ROC curve hugs the top left corner that results in an $AUC$ of 1. Fig. \ref{fig:ROC} depicts the ROC curves for all the 24 network topologies. As noticed, the ROC curves show that DA performs near to ideal for this problem, such that all the curves are close to the top left corner, even for the farthest curve associated with the C5-11 topology ($AUC$=0.995).

\indent After the training phase, DA is implemented on the test set. To access the performance of the proposed approach, three of the approaches reviewed in Section \ref{sec:introduction}, that have the same capabilities and applications as DA's, namely artificial neural network \cite{NN_topology_iden}, recursive Bayesian \cite{5424119}, and maximum likelihood \cite{6175638}, are implemented on this test feeder, as well. Table \ref{tab:Compar_orig} presents the misidentification rate of DA and its counterparts, for each switching configuration. In this table, to facilitate the comparison, the results are combined, such that topologies with the same switching configurations are considered together (for example the results of C1-00, C1-01, C1-10, and C1-11 are combined to form C1).

Considering the results provided in Table \ref{tab:Compar_orig}, while the artificial neural network approach is somehow successful in identifying the correct topology, but its misidentification rate is greater than that of DA, especially in regard to identifying the status of the protective devices. Furthermore, the 61.98\% misidentification rate concerning the weakly-meshed configuration, C2, shows that the recursive Bayesian approach fails to distinguish this configuration from the others. The superiority of DA over the recursive Bayesian approach can be justified by considering that unlike DA, the recursive Bayesian approach does not make use of the covariance of the measurements in its formulation, but only their variances. It can also be noticed that the performance of the maximum likelihood approach for this case study is poor. The main reason is that this method is based on linearizing the power flow equations around the mean of the pseudo-measurements. The problem is, the auto-regulators in this test feeder make the system's behavior absolutely nonlinear. In addition, pseudo-measurements generally contain a high level of uncertainty, which results in high deviations around their mean in the simulations. This has made the linearity assumption more inaccurate. In fact, when we decreased the standard deviation of the simulations from 0.3 to 0.03, the performance of the maximum likelihood approach was comparable to DA's, but this case would not present a practical distribution system.




\begin{table}[tp]
\centering
\caption{The performance of different TI approaches in identifying the correct network topology: the main case.}
\scalebox{0.7}{\addtolength{\tabcolsep}{-3pt}
\begin{tabular}{c|c|c|c|c|c|c|c|c|c}
\hline
\multicolumn{2}{c||}{Method$^{\mathrm{a}}$ $\rightarrow$} & 
\multicolumn{2}{c||}{DA} &
\multicolumn{2}{c||}{ANN \cite{NN_topology_iden}} &
\multicolumn{2}{c||}{RB \cite{5424119}} &
\multicolumn{2}{c}{ML \cite{6175638}}  \\
\hline
\multicolumn{2}{c||}{MI of$^{\mathrm{b}}$: $\rightarrow$} &
\multicolumn{1}{c|}{SC$^{\mathrm{c}}$} &
\multicolumn{1}{c||}{PDS$^{\mathrm{d}}$} &
\multicolumn{1}{c|}{SC} &
\multicolumn{1}{c||}{PDS} &
\multicolumn{1}{c|}{SC} &
\multicolumn{1}{c||}{PDS}&
\multicolumn{1}{c|}{SC} &
\multicolumn{1}{c}{PDS}   \\
\hline
\multirow{6}{*}{\rotatebox[origin=c]{90}{\parbox[c]{1.5 cm}{\centering Actual SC}}} &
\multicolumn{1}{c||}{ C1} & 
\multicolumn{1}{c|}{0.00\%} &
\multicolumn{1}{c||}{1.33\%} &
\multicolumn{1}{c|}{0.04\%} &
\multicolumn{1}{c||}{7.29\%} &
\multicolumn{1}{c|}{0.65\%} &
\multicolumn{1}{c||}{2.65\%} &
\multicolumn{1}{c|}{0.00\%} &
\multicolumn{1}{c}{33.91\%}  \\
\hhline{~~~~~~~~~~}
&
\multicolumn{1}{c||}{ C2} & 
\multicolumn{1}{c|}{0.38\%} &
\multicolumn{1}{c||}{3.55\%} &
\multicolumn{1}{c|}{0.71\%} &
\multicolumn{1}{c||}{5.71\%} &
\multicolumn{1}{c|}{61.98\%} &
\multicolumn{1}{c||}{1.53\%}  &
\multicolumn{1}{c|}{0.40\%} &
\multicolumn{1}{c}{28.44\%}  \\
\hhline{~~~~~~~~~~}
&
\multicolumn{1}{c||}{ C3} & 
\multicolumn{1}{c|}{0.00\%} &
\multicolumn{1}{c||}{0.20\%} &
\multicolumn{1}{c|}{0.00\%} &
\multicolumn{1}{c||}{2.96\%} &
\multicolumn{1}{c|}{0.00\%} &
\multicolumn{1}{c||}{0.00\%} &
\multicolumn{1}{c|}{2.80\%} &
\multicolumn{1}{c}{47.43\%}  \\
\hhline{~~~~~~~~~~}
&
\multicolumn{1}{c||}{ C4} & 
\multicolumn{1}{c|}{0.00\%} &
\multicolumn{1}{c||}{1.47\%} &
\multicolumn{1}{c|}{0.00\%} &
\multicolumn{1}{c||}{7.67\%} &
\multicolumn{1}{c|}{0.00\%} &
\multicolumn{1}{c||}{3.80\%} &
\multicolumn{1}{c|}{4.21\%} &
\multicolumn{1}{c}{45.87\%}  \\
\hhline{~~~~~~~~~~}
&
\multicolumn{1}{c||}{ C5} & 
\multicolumn{1}{c|}{0.00\%} &
\multicolumn{1}{c||}{0.27\%} &
\multicolumn{1}{c|}{0.00\%} &
\multicolumn{1}{c||}{4.88\%} &
\multicolumn{1}{c|}{0.02\%} &
\multicolumn{1}{c||}{2.82\%} &
\multicolumn{1}{c|}{21.65\%} &
\multicolumn{1}{c}{27.27\%}  \\
\hhline{~~~~~~~~~~}
&
\multicolumn{1}{c||}{ C6} & 
\multicolumn{1}{c|}{0.03\%} &
\multicolumn{1}{c||}{1.82\%} &
\multicolumn{1}{c|}{1.04\%} &
\multicolumn{1}{c||}{3.42\%} &
\multicolumn{1}{c|}{0.16\%} &
\multicolumn{1}{c||}{4.57\%} &
\multicolumn{1}{c|}{11.88\%} &
\multicolumn{1}{c}{40.40\%}  \\
\hline
\multicolumn{2}{c||}{Average} & 
\multicolumn{1}{c|}{0.07\%} &
\multicolumn{1}{c||}{1.44\%} &
\multicolumn{1}{c|}{0.30\%} &
\multicolumn{1}{c||}{5.32\%} &
\multicolumn{1}{c|}{10.47\%} &
\multicolumn{1}{c||}{2.66\%} &
\multicolumn{1}{c|}{6.82\%} &
\multicolumn{1}{c}{37.22\%}  \\
\hline
\multicolumn{10}{l}{$^{\mathrm{a}}$ANN = Artificial Neural Network,  RB = Recursive Bayesian,} \\
\hhline{~~~~~~~~~~} 
\multicolumn{10}{l}{ML = Maximum Likelihood;} \\
\hhline{~~~~~~~~~~} 
\multicolumn{10}{l}{$^{\mathrm{b}}$Misidentification of; $^{\mathrm{c}}$Switching Configuration; $^{\mathrm{d}}$Protective Devices Status.} \\
\hhline{~~~~~~~~~~}
\end{tabular}} \label{tab:Compar_orig}
\end{table}


\subsection{Recovery of Missing Signals to Enhance Resiliency}

\begin{figure}[tp]
\centering
\includegraphics[width=\linewidth]{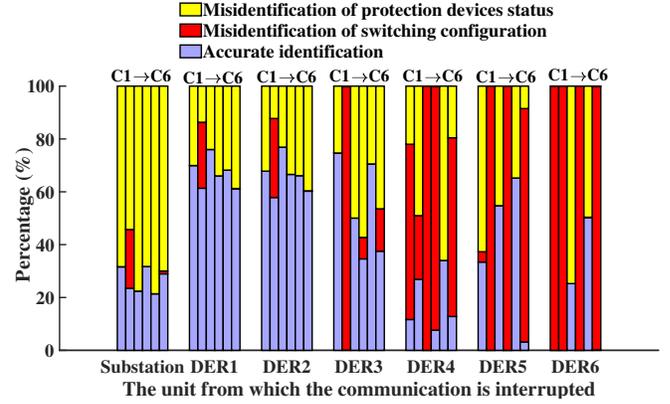}
\caption{The success rate of DA in TI, subjected to loss of the communication from one unit: considering the mean values in the training set for the missing signals.}
\label{fig:no_commu}
\end{figure}

\begin{table}[tp]
\centering
\caption{Correlation coefficients between the recovered and actual values of the missing signals.}
\scalebox{0.8}{\addtolength{\tabcolsep}{-2pt}
\begin{tabular}{c|c|c|c|c|c|c|c|c|c}
\hline
\multicolumn{1}{c||}{} &
\multicolumn{7}{c}{ The unit from which the communication is interrupted}\\ 
\hhline{~-------}
\multicolumn{1}{c||}{ Signal} &
\multicolumn{1}{c|}{\ Sub.} &
\multicolumn{1}{c|}{ DER1} &
\multicolumn{1}{c|}{ DER2} &
\multicolumn{1}{c|}{ DER3} &
\multicolumn{1}{c|}{ DER4} &
\multicolumn{1}{c|}{ DER5} &
\multicolumn{1}{c}{ DER6}  \\
\hline
\multicolumn{1}{c||}{ $v^+$} & 
\multicolumn{1}{c|}{ 0.03} &
\multicolumn{1}{c|}{ 0.52} &
\multicolumn{1}{c|}{ 0.51} &
\multicolumn{1}{c|}{ 0.99} &
\multicolumn{1}{c|}{ 0.98} &
\multicolumn{1}{c|}{ 0.99} &
\multicolumn{1}{c}{ 0.99} \\
\hhline{~~~~~~~~~~}
\multicolumn{1}{c||}{ $v^-$} & 
\multicolumn{1}{c|}{ 0.98} &
\multicolumn{1}{c|}{ 0.98} &
\multicolumn{1}{c|}{ 0.98} &
\multicolumn{1}{c|}{ 0.99} &
\multicolumn{1}{c|}{ 1} &
\multicolumn{1}{c|}{ 0.98} &
\multicolumn{1}{c}{ 1} \\
\hhline{~~~~~~~~~~}
\multicolumn{1}{c||}{ $P$} & 
\multicolumn{1}{c|}{ 0.93} &
\multicolumn{1}{c|}{ 0.84} &
\multicolumn{1}{c|}{ 0.84} &
\multicolumn{1}{c|}{ 0.78} &
\multicolumn{1}{c|}{ 0.82} &
\multicolumn{1}{c|}{ 0.83} &
\multicolumn{1}{c}{ 0.82} \\
\hhline{~~~~~~~~~~}
\multicolumn{1}{c||}{ $Q$} & 
\multicolumn{1}{c|}{ 0.95} &
\multicolumn{6}{c}{} \\
\hline
\end{tabular}} \label{tab:corr}
\end{table}

\begin{figure}[tp]
\centering
\includegraphics[width=\linewidth]{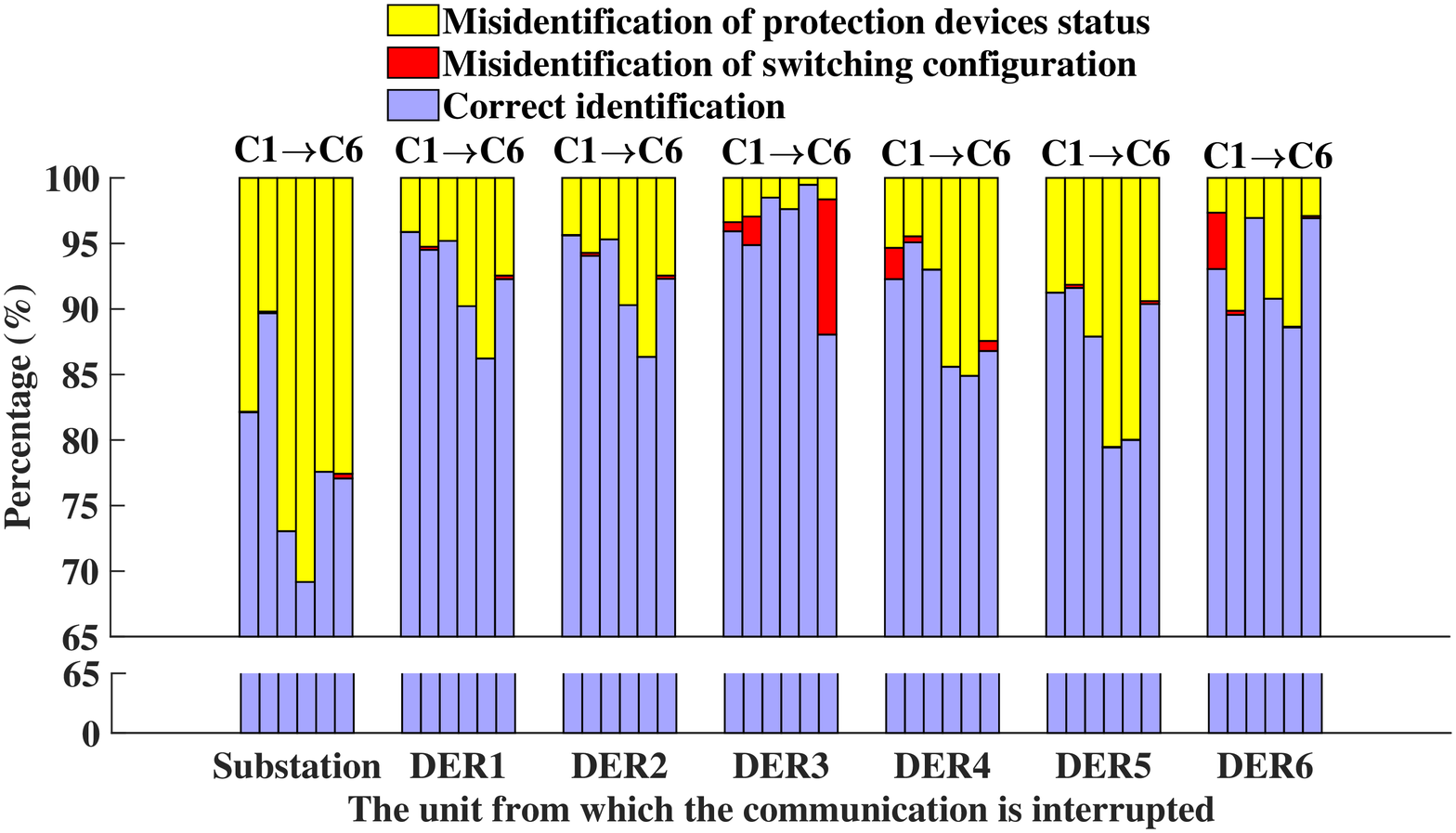}
\caption{The success rate of DA in TI, subjected to the loss of the communication from one unit: using the recovered signals.}
\label{fig:recov}
\end{figure}

\begin{figure}[tp]
\centering
\includegraphics[width=\linewidth]{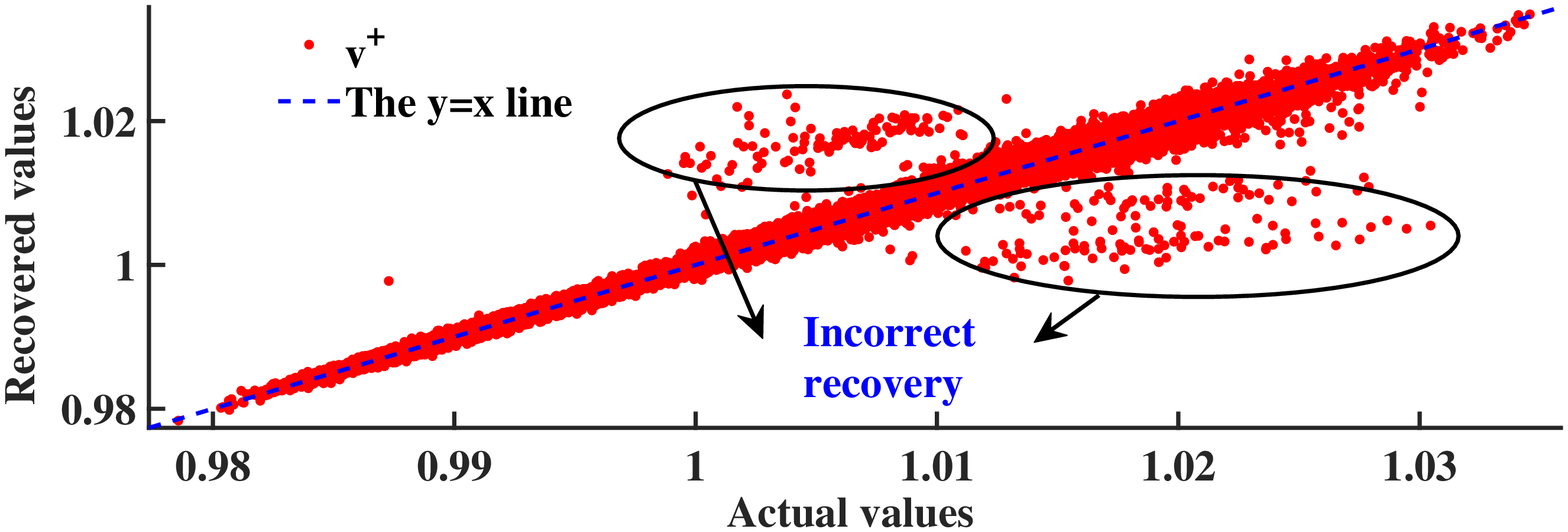}
\caption{The recovered versus actual values of $V^+$ for the case that the DER3 communication is interrupted.}
\label{fig:vplus_DER3}
\end{figure}

\begin{figure}[tp]
\centering
\includegraphics[width=\linewidth]{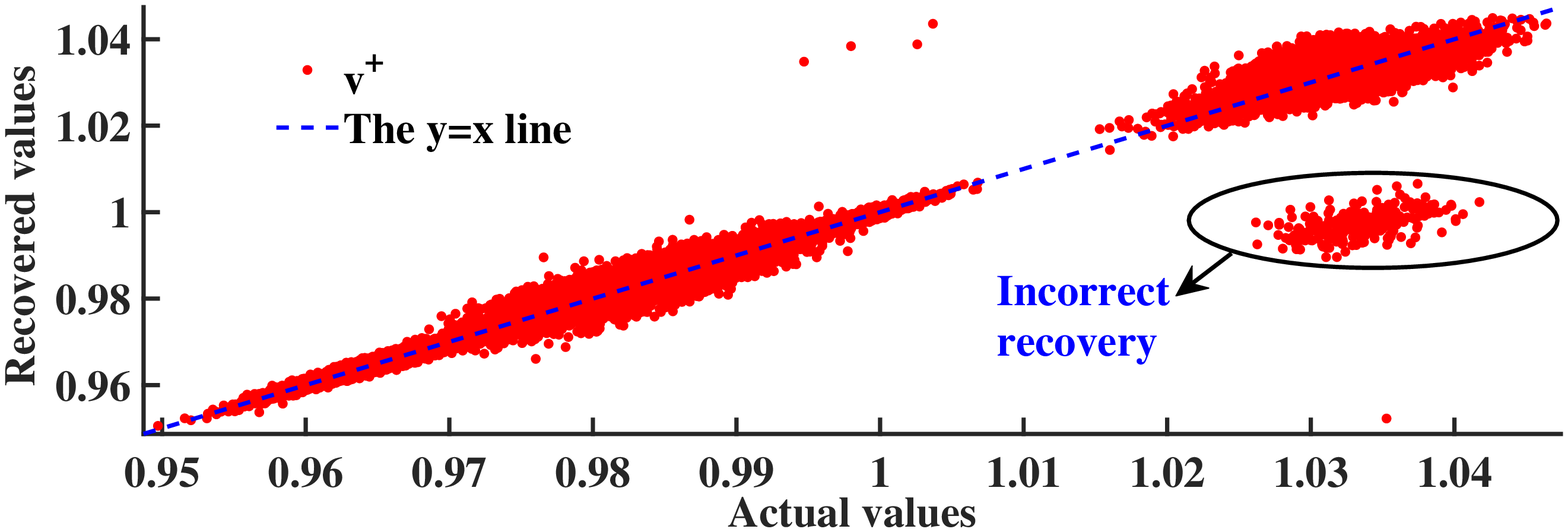}
\caption{The recovered versus actual values of $V^+$ for the case that the DER6 communication is interrupted.}
\label{fig:vplus_DER6}
\end{figure}

\begin{table}[tp]
\centering
\caption{The performance of different TI approaches in identifying the correct network topology: loss of the communication from DER3.}
\scalebox{0.7}{\addtolength{\tabcolsep}{-3pt}
\begin{tabular}{c|c|c|c|c|c|c|c|c|c}
\hline
\multicolumn{2}{c||}{Method$^{\mathrm{a}}$ $\rightarrow$} & 
\multicolumn{2}{c||}{Resilient DA} &
\multicolumn{2}{c||}{ANN \cite{NN_topology_iden}} &
\multicolumn{2}{c||}{RB \cite{5424119}} &
\multicolumn{2}{c}{ML \cite{6175638}}  \\
\hline
\multicolumn{2}{c||}{MI of$^{\mathrm{b}}$: $\rightarrow$} &
\multicolumn{1}{c|}{SC$^{\mathrm{c}}$} &
\multicolumn{1}{c||}{PDS$^{\mathrm{d}}$} &
\multicolumn{1}{c|}{SC} &
\multicolumn{1}{c||}{PDS} &
\multicolumn{1}{c|}{SC} &
\multicolumn{1}{c||}{PDS}&
\multicolumn{1}{c|}{SC} &
\multicolumn{1}{c}{PDS}   \\
\hline
\multirow{6}{*}{\rotatebox[origin=c]{90}{\parbox[c]{1.5 cm}{\centering Actual SC}}} &
\multicolumn{1}{c||}{ C1} & 
\multicolumn{1}{c|}{0.70\%} &
\multicolumn{1}{c||}{3.37\%} &
\multicolumn{1}{c|}{3.88\%} &
\multicolumn{1}{c||}{15.58\%} &
\multicolumn{1}{c|}{3.25\%} &
\multicolumn{1}{c||}{1.14\%}  &
\multicolumn{1}{c|}{0.00\%} &
\multicolumn{1}{c}{38.63\%}  \\
\hhline{~~~~~~~~~~}
&
\multicolumn{1}{c||}{ C2} & 
\multicolumn{1}{c|}{2.17\%} &
\multicolumn{1}{c||}{2.95\%} &
\multicolumn{1}{c|}{98.25\%} &
\multicolumn{1}{c||}{0.04\%} &
\multicolumn{1}{c|}{40.45\%} &
\multicolumn{1}{c||}{1.34\%}  &
\multicolumn{1}{c|}{0.08\%} &
\multicolumn{1}{c}{43.43\%}  \\
\hhline{~~~~~~~~~~}
&
\multicolumn{1}{c||}{ C3} & 
\multicolumn{1}{c|}{0.00\%} &
\multicolumn{1}{c||}{1.50\%} &
\multicolumn{1}{c|}{11.13\%} &
\multicolumn{1}{c||}{28.58\%} &
\multicolumn{1}{c|}{0.00\%} &
\multicolumn{1}{c||}{0.45\%} &
\multicolumn{1}{c|}{0.35\%} &
\multicolumn{1}{c}{49.87\%}  \\
\hhline{~~~~~~~~~~}
&
\multicolumn{1}{c||}{ C4} & 
\multicolumn{1}{c|}{0.00\%} &
\multicolumn{1}{c||}{2.37\%} &
\multicolumn{1}{c|}{0.08\%} &
\multicolumn{1}{c||}{27.92\%} &
\multicolumn{1}{c|}{0.00\%} &
\multicolumn{1}{c||}{4.05\%} &
\multicolumn{1}{c|}{1.98\%} &
\multicolumn{1}{c}{48.32\%}  \\
\hhline{~~~~~~~~~~}
&
\multicolumn{1}{c||}{ C5} & 
\multicolumn{1}{c|}{0.00\%} &
\multicolumn{1}{c||}{0.52\%} &
\multicolumn{1}{c|}{2.46\%} &
\multicolumn{1}{c||}{17.46\%} &
\multicolumn{1}{c|}{10.33\%} &
\multicolumn{1}{c||}{1.86\%} &
\multicolumn{1}{c|}{11.00\%} &
\multicolumn{1}{c}{40.87\%}  \\
\hhline{~~~~~~~~~~}
&
\multicolumn{1}{c||}{ C6} & 
\multicolumn{1}{c|}{10.32\%} &
\multicolumn{1}{c||}{1.63\%} &
\multicolumn{1}{c|}{77.17\%} &
\multicolumn{1}{c||}{1.67\%} &
\multicolumn{1}{c|}{3.94\%} &
\multicolumn{1}{c||}{3.39\%} &
\multicolumn{1}{c|}{52.92\%} &
\multicolumn{1}{c}{27.05\%}  \\
\hline
\multicolumn{2}{c||}{Average} & 
\multicolumn{1}{c|}{2.20\%} &
\multicolumn{1}{c||}{2.06\%} &
\multicolumn{1}{c|}{32.16\%} &
\multicolumn{1}{c||}{15.21\%} &
\multicolumn{1}{c|}{9.66\%} &
\multicolumn{1}{c||}{2.04\%} &
\multicolumn{1}{c|}{11.06\%} &
\multicolumn{1}{c}{41.36\%}  \\
\hline
\multicolumn{10}{l}{$^{\mathrm{a}}$ANN = Artificial Neural Network,  RB = Recursive Bayesian,} \\
\hhline{~~~~~~~~~~} 
\multicolumn{10}{l}{ML = Maximum Likelihood;} \\
\hhline{~~~~~~~~~~} 
\multicolumn{10}{l}{$^{\mathrm{b}}$Misidentification of; $^{\mathrm{c}}$Switching Configuration; $^{\mathrm{d}}$Protective Devices Status.} \\
\hhline{~~~~~~~~~~}
\end{tabular}} \label{tab:Compar_loss_DER3}
\end{table}

To examine the performance of the proposed data recovery approach, the interruption of the communication channels from the substation or one of the DERs is simulated, resulting in simultaneous loss of four signals in the former case or three signals in the latter cases. 
Without a recovery plan, considering the mean values in the training set for the missing signals seems like the most logical choice. Fig. \ref{fig:no_commu} presents the performance of DA in this scenario. As noted, DA can no longer predict the correct configuration, especially for the cases in which the communication from DER4, DER5, or DER6 is interrupted. This establishes the vulnerability of DA under the interruption of communication channels.

\indent Afterward, \eqref{eq:quadprog} is employed to recover the values of the missing signals. Table \ref{tab:corr} shows the correlation coefficients between the recovered and actual values of the missing signals. 
It can be noticed that the proposed approach has recovered $v^-$ with a high correlation to the actual values. The correlation coefficients are also high for $P$ and $Q$. However, the correlation is relatively low regarding $V^+$ for the cases that the communication from DER1 or DER2 is interrupted. The main reason for that is that these two DERs are electrically close to each other and therefore, there exists a high correlation between their voltage values (99.79\%), and hence, they are redundant regarding the classification process. Another point is that this approach has failed to recover the values of $V^+$ for the case in which the communication from the substation is interrupted. This is mainly because the transformer at the substation controls the $V^+$ of the substation MV-side (and not the $V^-$) and therefore, its contribution to the classification process is insignificant.

\indent Fig. \ref{fig:recov} presents the performance of DA with the recovered signals. Comparing to the results shown in Fig. \ref{fig:no_commu}, great enhancement in the performance of DA is noticeable, especially in the recognition of the correct switching configuration. The only exceptions are C6 and C1, for the cases in which communication from DER3 and DER6 is interrupted, respectively. Considering the recovered versus actual values of $V^+$ for these cases in Figs. \ref{fig:vplus_DER3} and \ref{fig:vplus_DER6}, respectively, it can be noticed that for a small portion of the observations, this approach has recovered incorrect values (C2 instead of C6 in the former and C2 instead of C1 in the latter case) and that is the main reason for the incorrect identification of the switching configuration in these cases. This implies that hardening the communication from DER6 and DER3 has a higher priority, regarding the resilient identification of the switching configuration.

\indent It should be remarked that instead of recovering the missing signals, an alternative idea is to build a model bank, in which each model is trained without measurements from one of the units, and these models would be employed in the case of the communication interruption.
However, this does not provide the ability to detect anomalous measurements, nor is it applicable if, e.g., only one signal is not available.

As, according to the results provided in Fig. \ref{fig:recov}, the least accuracy of DA in identifying the correct switching configuration is concerned with the DER3 communication interruption, this case is selected to compare the performance of DA with the performance of its counterparts. Table \ref{tab:Compar_loss_DER3} presents the results. It can be seen that under this scenario, the performance of the artificial neural network approach has greatly deteriorated, which highlights its susceptibility. The performance of the other two approaches is still not acceptable.

\subsection{Detecting Anomalous Measurements}

\begin{figure}[tp]
\centering
\includegraphics[width=\linewidth]{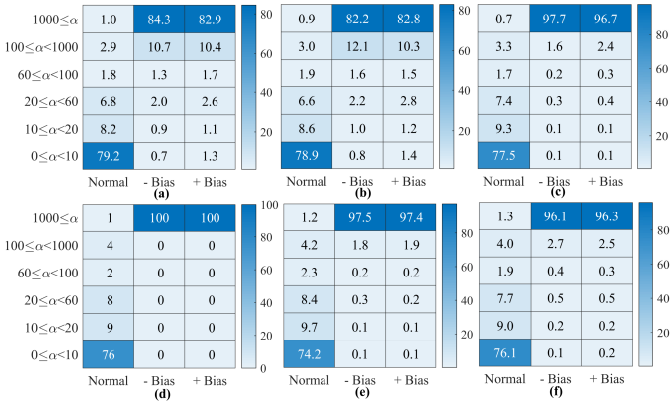}
\caption{The percentage of the data points in the test set placed in each specified range of $\alpha$, with the normal values and manipulated values of the measurements from: a) DER1, b) DER2, c) DER3, d) DER4, e) DER5, f) DER6.}
\label{fig:heatmap}
\end{figure}

\begin{table}[tp]
\centering
\caption{The performance of different TI approaches in identifying the correct network topology: the DER3 measurements contain anomalous values.}
\scalebox{0.7}{\addtolength{\tabcolsep}{-3pt}
\begin{tabular}{c|c|c|c|c|c|c|c|c|c}
\hline
\multicolumn{2}{c||}{Method$^{\mathrm{a}}$ $\rightarrow$} & 
\multicolumn{2}{c||}{Resilient DA} &
\multicolumn{2}{c||}{ANN \cite{NN_topology_iden}} &
\multicolumn{2}{c||}{RB \cite{5424119}} &
\multicolumn{2}{c}{ML \cite{6175638}}  \\
\hline
\multicolumn{2}{c||}{MI of$^{\mathrm{d}}$: $\rightarrow$} &
\multicolumn{1}{c|}{SC$^{\mathrm{e}}$} &
\multicolumn{1}{c||}{PDS$^{\mathrm{f}}$} &
\multicolumn{1}{c|}{SC} &
\multicolumn{1}{c||}{PDS} &
\multicolumn{1}{c|}{SC} &
\multicolumn{1}{c||}{PDS}&
\multicolumn{1}{c|}{SC} &
\multicolumn{1}{c}{PDS}   \\
\hline

\multirow{6}{*}{\rotatebox[origin=c]{90}{\parbox[c]{1.5 cm}{\centering Actual SC}}} &
\multicolumn{1}{c||}{ C1} & 
\multicolumn{1}{c|}{0.70\%} &
\multicolumn{1}{c||}{3.37\%} &
\multicolumn{1}{c|}{12.75\%} &
\multicolumn{1}{c||}{18.08\%} &
\multicolumn{1}{c|}{100\%} &
\multicolumn{1}{c||}{0.00\%}  &
\multicolumn{1}{c|}{0.00\%} &
\multicolumn{1}{c}{75.00\%}  \\
\hhline{~~~~~~~~~~}
&
\multicolumn{1}{c||}{ C2} & 
\multicolumn{1}{c|}{2.60\%} &
\multicolumn{1}{c||}{2.83\%} &
\multicolumn{1}{c|}{96.21\%} &
\multicolumn{1}{c||}{0.83\%} &
\multicolumn{1}{c|}{100\%} &
\multicolumn{1}{c||}{0.00\%} &
\multicolumn{1}{c|}{100\%} &
\multicolumn{1}{c}{0.00\%}  \\
\hhline{~~~~~~~~~~}
&
\multicolumn{1}{c||}{ C3} & 
\multicolumn{1}{c|}{0.00\%} &
\multicolumn{1}{c||}{1.50\%} &
\multicolumn{1}{c|}{0.00\%} &
\multicolumn{1}{c||}{7.54\%} &
\multicolumn{1}{c|}{0.00\%} &
\multicolumn{1}{c||}{75.00\%} &
\multicolumn{1}{c|}{100\%} &
\multicolumn{1}{c}{0.00\%}  \\
\hhline{~~~~~~~~~~}
&
\multicolumn{1}{c||}{ C4} & 
\multicolumn{1}{c|}{0.00\%} &
\multicolumn{1}{c||}{2.37\%} &
\multicolumn{1}{c|}{0.00\%} &
\multicolumn{1}{c||}{10.88\%} &
\multicolumn{1}{c|}{100\%} &
\multicolumn{1}{c||}{0.00\%}  &
\multicolumn{1}{c|}{100\%} &
\multicolumn{1}{c}{0.00\%}  \\
\hhline{~~~~~~~~~~}
&
\multicolumn{1}{c||}{ C5} & 
\multicolumn{1}{c|}{0.00\%} &
\multicolumn{1}{c||}{0.52\%} &
\multicolumn{1}{c|}{0.25\%} &
\multicolumn{1}{c||}{12.00\%} &
\multicolumn{1}{c|}{100\%} &
\multicolumn{1}{c||}{0.00\%} &
\multicolumn{1}{c|}{100\%} &
\multicolumn{1}{c}{0.00\%}  \\
\hhline{~~~~~~~~~~}
&
\multicolumn{1}{c||}{ C6} & 
\multicolumn{1}{c|}{10.32\%} &
\multicolumn{1}{c||}{1.63\%} &
\multicolumn{1}{c|}{0.00\%} &
\multicolumn{1}{c||}{4.63\%} &
\multicolumn{1}{c|}{0.00\%} &
\multicolumn{1}{c||}{72.78\%} &
\multicolumn{1}{c|}{100\%} &
\multicolumn{1}{c}{0.00\%}  \\
\hline
\multicolumn{2}{c||}{Average} & 
\multicolumn{1}{c|}{2.27\%} &
\multicolumn{1}{c||}{2.04\%} &
\multicolumn{1}{c|}{18.20\%} &
\multicolumn{1}{c||}{8.99\%} &
\multicolumn{1}{c|}{66.67\%} &
\multicolumn{1}{c||}{24.63\%} &
\multicolumn{1}{c|}{83.33\%} &
\multicolumn{1}{c}{12.50\%}  \\
\hline
\multicolumn{10}{l}{$^{\mathrm{a}}$ANN = Artificial Neural Network,  RB = Recursive Bayesian,} \\
\hhline{~~~~~~~~~~} 
\multicolumn{10}{l}{ML = Maximum Likelihood;} \\
\hhline{~~~~~~~~~~} 
\multicolumn{10}{l}{$^{\mathrm{b}}$Misidentification of; $^{\mathrm{c}}$Switching Configuration; $^{\mathrm{d}}$Protective Devices Status.} \\
\hhline{~~~~~~~~~~}
\end{tabular}} \label{tab:Compar_anomaly_DER3}
\end{table}

To simulate anomalous measurements, at each step, the measurements from one of the DERs are multiplied by 0.9, to examine a negative bias, and then 1.1, to examine a positive bias. $\alpha$ is calculated with the manipulated values, as well as the original values.
Fig. \ref{fig:heatmap} presents the percentage of the data points placed in each specified range of $\alpha$. 
As noted, with the original values, for the majority of the data points $\alpha$ is below 60 (more than 92\% of the data points, in all the scenarios), while with the manipulated values, $\alpha$ takes larger values (above 60 for over 95\% of the data points, in all the scenarios), showing that $\alpha$ is an effective benchmark to detect anomalous measurements.

Table \ref{tab:Compar_anomaly_DER3} compares the performance of the proposed approach with the performance of its counterparts in the case that the DER3 measurements contain manipulated values. It can be noted that the performance of DA is similar to the case that DER3 communication is interrupted. This is because with the use of the proposed benchmark, $\alpha$, the manipulated measurements are successfully identified and replaced with the recovered values. On the contrary, the performance of the recursive Bayesian and maximum likelihood approaches is greatly influenced, such that they no longer can identify the correct switching configuration. The artificial neural network approach also does not provide an accurate solution either. In fact, this was expected as these approaches are not resilience-oriented.

\section{Sensitivity Analysis} \label{sec:sensitivity}
In this section, firstly the performance of the proposed approach is investigated assuming different configurations of DERs in the network, and then, the impact of the load type and loading level is studied. 

\subsection{Assessing The Contribution of DERs} 

\begin{figure}[tp]
\centering
\includegraphics[width=\linewidth]{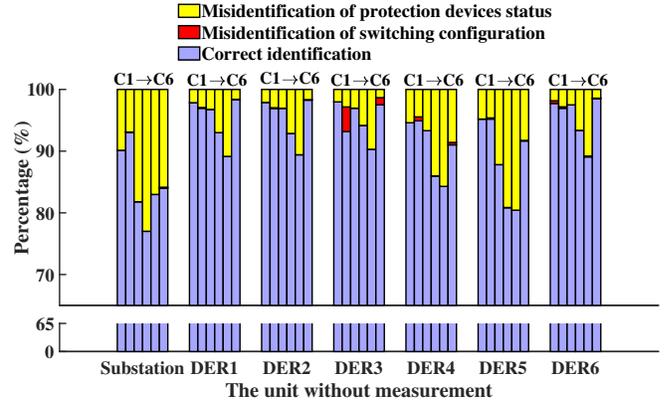}
\caption{The success rate of DA in TI: DA was trained and tested without the measurements from one of the units.}
\label{fig:loss_of_one}
\end{figure}

\begin{figure*}[htp]
\centering
\includegraphics[width=\linewidth]{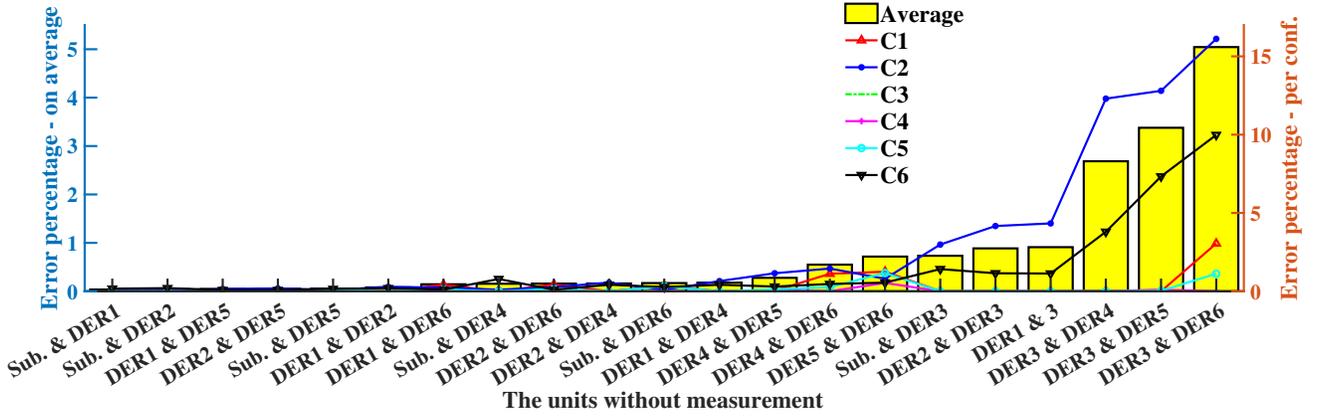}
\caption{The success rate of DA in TI: DA was trained and tested without the measurements from two units; the percentage of the switching configuration misclassification has been presented on average in the left axis and per configuration in the right axis.}
\label{fig:loss_of_two}
\end{figure*}

\begin{table}[tp]
\centering
\caption{Selected columns of confusion matrices for the cases that DA was trained and tested without the measurements from two units.}
\scalebox{0.7}{
\begin{tabular}{c|c|c|c|c|c|c|c}
\hline
\multicolumn{2}{c||}{Units without measure. $\rightarrow$} & 
\multicolumn{2}{c||}{DER3 \& DER6} &
\multicolumn{2}{c||}{DER3 \& DER5} &
\multicolumn{2}{c}{DER3 \& DER4}  \\
\hline
\multicolumn{2}{c||}{Correct sw. config. $\rightarrow$} &
\multicolumn{1}{c|}{ C2} &
\multicolumn{1}{c||}{ C6} &
\multicolumn{1}{c|}{ C2} &
\multicolumn{1}{c||}{ C6} &
\multicolumn{1}{c|}{ C2} &
\multicolumn{1}{c}{ C6}   \\
\hline
\multirow{6}{*}{\rotatebox[origin=c]{90}{\parbox[c]{1.5 cm}{\centering Predicted switching configuration}}} &
\multicolumn{1}{c||}{ C1} & 
\multicolumn{1}{c|}{ 0} &
\multicolumn{1}{c||}{ 0\%} &
\multicolumn{1}{c|}{ 0\%} &
\multicolumn{1}{c||}{ 0\%} &
\multicolumn{1}{c|}{ 0} &
\multicolumn{1}{c}{ 0\%}  \\
\hhline{~~~~~~~~}
&
\multicolumn{1}{c||}{ C2} & 
\multicolumn{1}{c|}{ 84\%} &
\multicolumn{1}{c||}{ \color[HTML]{FF0000} 10\%} &
\multicolumn{1}{c|}{ 87\%} &
\multicolumn{1}{c||}{ \color[HTML]{FF0000} 7\%} &
\multicolumn{1}{c|}{ 87\%} &
\multicolumn{1}{c}{ \color[HTML]{FF0000} 7\%}  \\
\hhline{~~~~~~~~}
&
\multicolumn{1}{c||}{ C3} & 
\multicolumn{1}{c|}{ 0} &
\multicolumn{1}{c||}{ 0} &
\multicolumn{1}{c|}{ 0} &
\multicolumn{1}{c||}{ 0} &
\multicolumn{1}{c|}{ 0} &
\multicolumn{1}{c}{ 0}   \\
\hhline{~~~~~~~~}
&
\multicolumn{1}{c||}{ C4} & 
\multicolumn{1}{c|}{ 0} &
\multicolumn{1}{c||}{ 0} &
\multicolumn{1}{c|}{ 0} &
\multicolumn{1}{c||}{ 0} &
\multicolumn{1}{c|}{ 0} &
\multicolumn{1}{c}{ 0}  \\
\hhline{~~~~~~~~}
&
\multicolumn{1}{c||}{ C5} & 
\multicolumn{1}{c|}{ 0} &
\multicolumn{1}{c||}{ 0} &
\multicolumn{1}{c|}{ 0} &
\multicolumn{1}{c||}{ 0} &
\multicolumn{1}{c|}{ 0} &
\multicolumn{1}{c}{ 0}  \\
\hhline{~~~~~~~~}
&
\multicolumn{1}{c||}{ C6} & 
\multicolumn{1}{c|}{ \color[HTML]{FF0000} 16\%} &
\multicolumn{1}{c||}{ 90\%} &
\multicolumn{1}{c|}{ \color[HTML]{FF0000} 13\%} &
\multicolumn{1}{c||}{ 93\%} &
\multicolumn{1}{c|}{ \color[HTML]{FF0000} 13\%} &
\multicolumn{1}{c}{ 93\%} \\
\hline
\multicolumn{2}{c||}{ Sum} & 
\multicolumn{1}{c|}{ 100\%} &
\multicolumn{1}{c||}{ 100\%} &
\multicolumn{1}{c|}{ 100\%} &
\multicolumn{1}{c||}{ 100\%} &
\multicolumn{1}{c|}{ 100\%} &
\multicolumn{1}{c}{ 100\%}  \\
\hline

\end{tabular}} \label{tab:Confusion_Matrix}
\end{table}

To investigate how the number and configuration of measurement points would influence the performance of the proposed topology identification method, firstly the model is trained and tested without the measurements from one of the units (substation or DERs). Fig. \ref{fig:loss_of_one} presents the results.
It is noted that regarding the identification of the network switching configuration, the method is still efficacious, while regarding the identification of the status of the protective devices, the performance of the model decreases to some extent in all the switching configurations without the measurements from the substation or DER5. It can be inferred that having a measurement unit in these areas has a high contribution to the identification of the status of the protective devices.

\indent Afterward, the model is trained and tested without the measurements from two of the units. Fig. \ref{fig:loss_of_two} presents the percentage of switching configuration misclassification on average in ascending order (the left axis) and per configuration (the right axis).  
Confusion matrix is often used to interpret the results of classification models. This matrix summarizes the percentage of correct and incorrect predictions, broken down by each class \cite{james2013introduction}. Table \ref{tab:Confusion_Matrix} presents selected columns of the confusion matrices for this case. 
Considering Fig. \ref{fig:loss_of_two} and Table \ref{tab:Confusion_Matrix} together, it can be noticed that almost in each combination that contains DER3, the misclassification percentage for C6 and C2 increases significantly, especially when DER3 is paired with DER6, DER5, and DER4. 
It can also be recognized that this occurs mainly because the model mixes up these two configurations. The only difference between these two configurations is the status of $S_{18-135}$ (open in C6 and close in C2). This implies that to perceive the status of $S_{18-135}$, it is important to have a measurement point in the zone between $S_{18-135}$ and $S_{151-300}$.

\subsection{Impact of Load Type And Loading Level}

\begin{table}[tp]
\centering
\caption{The performance of DA in identifying the correct network topology: the impact of loads type and loading level.}
\scalebox{0.7}{\addtolength{\tabcolsep}{-2pt}
\begin{tabular}{c|c|c|c|c|c|c|c}
\hline
\multicolumn{2}{c||}{Scenario $\rightarrow$} & 
\multicolumn{2}{c||}{Main} &
\multicolumn{2}{c||}{Loss of commu.} &
\multicolumn{2}{c}{DER3 anomalous}  \\
\hhline{~~~~~~~~}
\multicolumn{2}{c||}{} & 
\multicolumn{2}{c||}{} &
\multicolumn{2}{c||}{from DER3} &
\multicolumn{2}{c}{measurements}  \\
\hline
\multicolumn{2}{c||}{Misidentification of: $\rightarrow$} &
\multicolumn{1}{c|}{SC$^{\mathrm{a}}$} &
\multicolumn{1}{c||}{PDS$^{\mathrm{b}}$} &
\multicolumn{1}{c|}{SC} &
\multicolumn{1}{c||}{PDS} &
\multicolumn{1}{c|}{SC} &
\multicolumn{1}{c}{PDS}   \\
\hhline{--~~~~~~}

\multicolumn{1}{c|}{Load type $\downarrow$} &
\multicolumn{1}{c||}{Loading $\downarrow$} &
\multicolumn{1}{c|}{} &
\multicolumn{1}{c||}{} &
\multicolumn{1}{c|}{} &
\multicolumn{1}{c||}{} &
\multicolumn{1}{c|}{} &
\multicolumn{1}{c}{}   \\
\hline
\multicolumn{1}{c|}{Original} &
\multicolumn{1}{c||}{Original} & 
\multicolumn{1}{c|}{0.07\%} &
\multicolumn{1}{c||}{1.44\%} &
\multicolumn{1}{c|}{2.20\%} &
\multicolumn{1}{c||}{2.06\%} &
\multicolumn{1}{c|}{2.27\%} &
\multicolumn{1}{c}{2.04\%}  \\
\hhline{~~~~~~~~}
\multicolumn{1}{c|}{PQ$^{\mathrm{c}}$} &
\multicolumn{1}{c||}{Original} & 
\multicolumn{1}{c|}{0.10\%} &
\multicolumn{1}{c||}{1.70\%} &
\multicolumn{1}{c|}{1.70\%} &
\multicolumn{1}{c||}{2.37\%} &
\multicolumn{1}{c|}{1.72\%} &
\multicolumn{1}{c}{2.37\%}  \\
\hhline{~~~~~~~~}
\multicolumn{1}{c|}{Z$^{\mathrm{d}}$} &
\multicolumn{1}{c||}{Original} & 
\multicolumn{1}{c|}{0.02\%} &
\multicolumn{1}{c||}{3.50\%} &
\multicolumn{1}{c|}{1.61\%} &
\multicolumn{1}{c||}{3.40\%} &
\multicolumn{1}{c|}{1.61\%} &
\multicolumn{1}{c}{3.40\%}  \\
\hhline{~~~~~~~~}
\multicolumn{1}{c|}{I$^{\mathrm{e}}$} &
\multicolumn{1}{c||}{Original} & 
\multicolumn{1}{c|}{0.01\%} &
\multicolumn{1}{c||}{3.30\%} &
\multicolumn{1}{c|}{1.42\%} &
\multicolumn{1}{c||}{3.01\%} &
\multicolumn{1}{c|}{1.44\%} &
\multicolumn{1}{c}{3.01\%} \\
\hhline{~~~~~~~~}
\multicolumn{1}{c|}{Original} &
\multicolumn{1}{c||}{P 120\%} & 
\multicolumn{1}{c|}{0.02\%} &
\multicolumn{1}{c||}{0.32\%} &
\multicolumn{1}{c|}{1.18\%} &
\multicolumn{1}{c||}{0.48\%} &
\multicolumn{1}{c|}{1.21\%} &
\multicolumn{1}{c}{0.48\%}  \\
\hhline{~~~~~~~~}
\multicolumn{1}{c|}{Original} &
\multicolumn{1}{c||}{Q 120\%} & 
\multicolumn{1}{c|}{0.02\%} &
\multicolumn{1}{c||}{1.50\%} &
\multicolumn{1}{c|}{1.23\%} &
\multicolumn{1}{c||}{2.10\%} &
\multicolumn{1}{c|}{1.25\%} &
\multicolumn{1}{c}{2.10\%}   \\
\hline
\multicolumn{8}{l}{$^{\mathrm{a}}$ Switching configuration; $^{\mathrm{b}}$ Protective devices status;} \\
\hhline{~~~~~~~~} 
\multicolumn{8}{l}{$^{\mathrm{c}}$ Constant power; $^{\mathrm{d}}$ Constant impedance; $^{\mathrm{e}}$ Constant current.} \\
\hhline{~~~~~~~~}
\end{tabular}} \label{tab:Compar_loading}
\end{table}

The IEEE 123 test feeder contains different load types. To investigate how the system load type will influence the performance of the proposed approach, at each step, all the system loads are replaced with loads of similar values, but of homogeneous type, first, constant power, then constant impedance, and finally, constant current. Three scenarios are investigated: the main scenario, the loss of the DER3 communication scenario, and the scenario that the DER3 measurements are manipulated.

Table \ref{tab:Compar_loading} presents the average misidentification rate over all the switching configurations for these scenarios. It can be noticed that with the constant impedance and current load types, the switching configuration misidentification rate has improved, while the rate of protective devices status misidentification has deteriorated to some extent. This can be justified by considering that with the constant impedance and constant current load types, the dependency between the active and reactive power consumption of the nodes with the voltage of the nodes increases, and therefore, with the change of the switching configuration, the voltages become more distinguishable, and since DA employs voltage measurements in this study system, its performance in identifying the correct switching configuration improves. On the other hand, since in these cases, a lower magnitude of the load is curtailed by the operation of the protective devices (as the voltage of the areas beyond the protective devices in this test system is lower than 1 per unit), it becomes more arduous to differentiate these topologies from the others.

Afterward, to investigate the impact of the loading level, at first all the active power values, and then the reactive power values, are multiplied by 1.2, and simulations are repeated. Table \ref{tab:Compar_loading} presents the results. As seen, with the increase in the loading level, the misidentification rate of the stitching configuration decreases. It is because when the loading level increase, the voltage of the end nodes will decrease, and hence, the switching configurations become more distinguishable. Another noticeable trend is that with the increase in the active power, the rate of protective devices status misidentification shrinks. This is mainly because in this case, with the operation of the protective devices, a higher amount of active power is curtailed and this makes it more straightforward to detect their operation. This is not the case for the reactive power, as the power factor of loads in this feeder is around 0.9 lag, and therefore, the reactive power variations are less tractable than those of the active power.

To sum up, it can be concluded that the changes caused in the performance of the proposed approach with the change in the load type and loading level are mainly due to the topologies become less or more distinguishable by nature, and they do not noticeably influence the functionality of the proposed approach.

\section{Conclusion}\label{sec:conclusion}
A DA classification approach was proposed to identify the real-time topology of distribution networks that relies only on the measurements available to DERMS. The network topology changes due to the network reconfiguration operations and faults. The proposed approach was implemented on a modified version of the IEEE 123 node test feeder. Results demonstrated the superiority of DA over its counterparts in identifying the network switching configuration, as well as the protective devices. This approach could be implemented in practice, as it requires only a mean vector and a covariance matrix to calculate the probability of an observation belonging to a specific network topology.

\indent Afterward, a quadratic programming optimization approach, based on the recovery of the lost signals, was presented to make the proposed TI approach
resilient against the interruption of communication channels. The specific optimization problem can be settled efficiently and a global extremum is guaranteed. Furthermore, by exploiting this data recovery approach, a benchmark was introduced to detect anomalous measurements. This benchmark can be employed to enhance the resiliency of the proposed TI against cyber-attacks.

Sensitivity analysis shows that the load type and loading level do not noticeably influence the functionality of the proposed approach, although they might influence its performance as they can change the nature of the topology identification problem by making the system topologies less or more distinguishable.


Bibliography


%





\ifCLASSOPTIONcaptionsoff
  \newpage
\fi





\bibliographystyle{IEEEtran}
\bibliography{IEEEabrv,Bibliography}

\begin{thebibliography}{10}
\providecommand{\url}[1]{#1}
\csname url@rmstyle\endcsname
\providecommand{\newblock}{\relax}
\providecommand{\bibinfo}[2]{#2}
\providecommand\BIBentrySTDinterwordspacing{\spaceskip=0pt\relax}
\providecommand\BIBentryALTinterwordstretchfactor{4}
\providecommand\BIBentryALTinterwordspacing{\spaceskip=\fontdimen2\font plus
\BIBentryALTinterwordstretchfactor\fontdimen3\font minus
  \fontdimen4\font\relax}
\providecommand\BIBforeignlanguage[2]{{%
\expandafter\ifx\csname l@#1\endcsname\relax
\typeout{** WARNING: IEEEtran.bst: No hyphenation pattern has been}%
\typeout{** loaded for the language `#1'. Using the pattern for}%
\typeout{** the default language instead.}%
\else
\language=\csname l@#1\endcsname
\fi
#2}}
\renewcommand\BIBentryALTinterwordstretchfactor{4}

\bibitem{Argonne}
J.~Wang, X.~Lu, and C.~Chen, ``Guidelines for implementing advanced
  distribution management systems,'' Argonne National Laboratory, Tech. Rep.
  ANL/ESD-15/15, 2015.

\bibitem{EPRI}
B.~Seal, J.~Simmins, and G.~Gray, ``Enterprise integration functions for
  distributed energy resources,'' Electric Power Research Institute, Tech. Rep.
  3002001249, 2013.

\bibitem{8585806}
S.~{Nowak}, N.~{Tehrani}, M.~S. {Metcalfe}, W.~{Eberle}, and L.~{Wang},
  ``Cloud-based derms test platform using real-time power system simulation,''
  in \emph{2018 IEEE Power Energy Society General Meeting (PESGM)}, Aug 2018,
  pp. 1--5.

\bibitem{EPRI-understanding}
B.~Seal, A.~Renjit, and B.~Deaver, ``Understanding derms,'' Electric Power
  Research Institute, Tech. Rep. 3002013049, 2018.

\bibitem{DERMS_JOOS}
G.~Joos, ``Standards for specification of microgrid controllers (ieee std
  2030.7/8) and derms (ieee p2030.11).''\hskip 1em plus 0.5em minus 0.4em\relax
  Invited presentation at the 8th International Conference on Integration of
  Renewable and Distributed Energy Resources, 2018.

\bibitem{6175638}
Y.~{Sharon}, A.~M. {Annaswamy}, A.~L. {Motto}, and A.~{Chakraborty}, ``Topology
  identification in distribution network with limited measurements,'' in
  \emph{2012 IEEE PES Innovative Smart Grid Technologies (ISGT)}, Jan 2012, pp.
  1--6.

\bibitem{5424119}
R.~{Singh}, E.~{Manitsas}, B.~C. {Pal}, and G.~{Strbac}, ``A recursive bayesian
  approach for identification of network configuration changes in distribution
  system state estimation,'' \emph{IEEE Transactions on Power Systems},
  vol.~25, no.~3, pp. 1329--1336, Aug 2010.

\bibitem{7027869}
Z.~{Tian}, W.~{Wu}, and B.~{Zhang}, ``A mixed integer quadratic programming
  model for topology identification in distribution network,'' \emph{IEEE
  Transactions on Power Systems}, vol.~31, no.~1, pp. 823--824, Jan 2016.

\bibitem{8754743}
G.~{Cavraro}, A.~{Bernstein}, V.~{Kekatos}, and Y.~{Zhang}, ``Real-time
  identifiability of power distribution network topologies with limited
  monitoring,'' \emph{IEEE Control Systems Letters}, vol.~4, no.~2, pp.
  325--330, 2020.

\bibitem{8787584}
M.~{Farajollahi}, A.~{Shahsavari}, and H.~{Mohsenian-Rad}, ``Topology
  identification in distribution systems using line current sensors: An milp
  approach,'' \emph{IEEE Transactions on Smart Grid}, pp. 1--1, 2019.

\bibitem{7856295}
B.~{Hayes}, A.~{Escalera}, and M.~{Prodanovic}, ``Event-triggered topology
  identification for state estimation in active distribution networks,'' in
  \emph{2016 IEEE PES Innovative Smart Grid Technologies Conference Europe
  (ISGT-Europe)}, Oct 2016, pp. 1--6.

\bibitem{1490609}
J.~R. {Aguero} and A.~{Vargas}, ``Inference of operative configuration of
  distribution networks using fuzzy logic techniques-part i: real-time model,''
  \emph{IEEE Transactions on Power Systems}, vol.~20, no.~3, pp. 1551--1561,
  Aug 2005.

\bibitem{NN_topology_iden}
M.~Jafarian, A.~Soroudi, and A.~Keane, ``Distribution system topology
  identification for der management systems using deep neural networks,'' in
  \emph{2020 IEEE Power \& Energy Society General Meeting}, August 2020.

\bibitem{4810103}
R.~{Singh}, B.~C. {Pal}, and R.~B. {Vinter}, ``Measurement placement in
  distribution system state estimation,'' \emph{IEEE Transactions on Power
  Systems}, vol.~24, no.~2, pp. 668--675, May 2009.

\bibitem{ARGHANDEH20161060}
R.~Arghandeh, A.~von Meier, L.~Mehrmanesh, and L.~Mili, ``On the definition of
  cyber-physical resilience in power systems,'' \emph{Renewable and Sustainable
  Energy Reviews}, vol.~58, pp. 1060 -- 1069, 2016.

\bibitem{hooshyar1}
A.~Ameli, A.~Hooshyar, A.~Yazdavar, E.~El-Saadany, and A.~Youssef, ``Attack
  detection for load frequency control systems using stochastic unknown input
  estimators,'' \emph{IEEE Transactions on Information Forensics and Security},
  vol.~PP, pp. 1--1, 04 2018.

\bibitem{hooshyar2}
A.~Ameli, A.~Hooshyar, E.~El-Saadany, and A.~Youssef, ``Attack detection and
  identification for automatic generation control systems,'' \emph{IEEE
  Transactions on Power Systems}, vol.~PP, pp. 1--1, 02 2018.

\bibitem{most_common}
J.~E. Holden, W.~H. Finch, and K.~Kelley, ``A comparison of two-group
  classification methods,'' \emph{Educational and Psychological Measurement},
  vol.~71, no.~5, pp. 870--901, 2011.

\bibitem{SERRANOCINCA20131245}
C.~Serrano-Cinca and B.~Gutiérrez-Nieto, ``Partial least square discriminant
  analysis for bankruptcy prediction,'' \emph{Decision Support Systems},
  vol.~54, no.~3, pp. 1245 -- 1255, 2013.

\bibitem{6508865}
W.~{Zhou}, Y.~{Liu}, Q.~{Yuan}, and X.~{Li}, ``Epileptic seizure detection
  using lacunarity and bayesian linear discriminant analysis in intracranial
  eeg,'' \emph{IEEE Transactions on Biomedical Engineering}, vol.~60, no.~12,
  pp. 3375--3381, Dec 2013.

\bibitem{CHIEN2018379}
N.~P. Chien and L.~K. Lautz, ``Discriminant analysis as a decision-making tool
  for geochemically fingerprinting sources of groundwater salinity,''
  \emph{Science of The Total Environment}, vol. 618, pp. 379 -- 387, 2018.

\bibitem{49084}
N.~F. {Hubele} and C.~. {Cheng}, ``Identification of seasonal short-term load
  forecasting models using statistical decision functions,'' \emph{IEEE
  Transactions on Power Systems}, vol.~5, no.~1, pp. 40--45, Feb 1990.

\bibitem{6476045}
Y.~{Li} and P.~J. {Wolfs}, ``A hybrid model for residential loads in a
  distribution system with high pv penetration,'' \emph{IEEE Transactions on
  Power Systems}, vol.~28, no.~3, pp. 3372--3379, Aug 2013.

\bibitem{7066231}
Y.~{Wang}, J.~{Zhou}, Z.~{Li}, Z.~{Dong}, and Y.~{Xu},
  ``Discriminant-analysis-based single-phase earth fault protection using
  improved pca in distribution systems,'' \emph{IEEE Transactions on Power
  Delivery}, vol.~30, no.~4, pp. 1974--1982, Aug 2015.

\bibitem{153398}
G.~{Garcia}, J.~{Bernussou}, and M.~{Berbiche}, ``Pattern recognition applied
  to transient stability analysis of power systems with modelling including
  voltage and speed regulation,'' \emph{IEE Proceedings B - Electric Power
  Applications}, vol. 139, no.~4, pp. 321--335, July 1992.

\bibitem{962423}
C.~A. {Jensen}, M.~A. {El-Sharkawi}, and R.~J. {Marks}, ``Power system security
  assessment using neural networks: feature selection using fisher
  discrimination,'' \emph{IEEE Transactions on Power Systems}, vol.~16, no.~4,
  pp. 757--763, Nov 2001.

\bibitem{James}
G.~M. James and T.~J. Hastie, ``Functional linear discriminant analysis for
  irregularly sampled curves,'' \emph{Journal of the Royal Statistical Society:
  Series B (Statistical Methodology)}, vol.~63, no.~3, pp. 533--550, 2001.

\bibitem{james2013introduction}
G.~James, D.~Witten, T.~Hastie, and R.~Tibshirani, \emph{An Introduction to
  Statistical Learning: with Applications in R}, ser. Springer Texts in
  Statistics.\hskip 1em plus 0.5em minus 0.4em\relax Springer New York, 2013.

\bibitem{38136}
K.~P. Murphy, \emph{Machine learning: a probabilistic perspective}, Cambridge,
  MA, 2012.

\bibitem{robust_1}
J.~Hallinan, ``Chapter 2 - data mining for microbiologists,'' in \emph{Systems
  Biology of Bacteria}, ser. Methods in Microbiology, C.~Harwood and A.~Wipat,
  Eds.\hskip 1em plus 0.5em minus 0.4em\relax Academic Press, 2012, vol.~39,
  pp. 27 -- 79.

\bibitem{robust_2}
W.~R. Dillon, ``The performance of the linear discriminant function in
  nonoptimal situations and the estimation of classification error rates: A
  review of recent findings,'' \emph{Journal of Marketing Research}, vol.~16,
  no.~3, pp. 370--381, 1979.

\bibitem{6204095}
M.~{Biserica}, Y.~{Besanger}, R.~{Caire}, O.~{Chilard}, and P.~{Deschamps},
  ``Neural networks to improve distribution state estimation—volt var control
  performances,'' \emph{IEEE Transactions on Smart Grid}, vol.~3, no.~3, pp.
  1137--1144, Sep. 2012.

\bibitem{IrelandGridCode}
``Distribution code,'' ESB Networks,'' Grid code, April 2016.

\bibitem{119237}
W.~H. {Kersting}, ``Radial distribution test feeders,'' \emph{IEEE Transactions
  on Power Systems}, vol.~6, no.~3, pp. 975--985, Aug 1991.

\bibitem{5419278}
C.~{Chang}, ``Multiparameter receiver operating characteristic analysis for
  signal detection and classification,'' \emph{IEEE Sensors Journal}, vol.~10,
  no.~3, pp. 423--442, March 2010.

\end{thebibliography}
%

\begin{IEEEbiography}[{\includegraphics[width=1in,height=1.25in,clip,keepaspectratio]{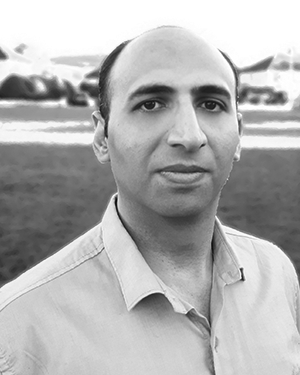}}]{Mohammad Jafarian}(Member, IEEE)
received the Ph.D. degree in electrical engineering from Sharif University of Technology, Tehran, Iran, in 2013. He is a senior power systems researcher with the Energy Institute at the University College Dublin. His research interests include power system dynamics, wind power integration, and distribution networks.
\end{IEEEbiography}
\begin{IEEEbiography}[{\includegraphics[width=1in,height=1.25in,clip,keepaspectratio]{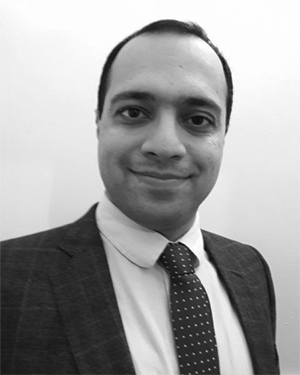}}]{Alireza Soroudi}(Senior Member, IEEE)
received the Ph.D. degree in electrical engineering from the Grenoble-INP, Grenoble, France, in 2012. He is currently an Assistant Professor with the Electrical and Electronic Engineering School at University College Dublin, Dublin, Ireland. His research interests include power systems planning and operation, risk, and uncertainty modelling.
\end{IEEEbiography}
\begin{IEEEbiography}[{\includegraphics[width=1in,height=1.25in,clip,keepaspectratio]{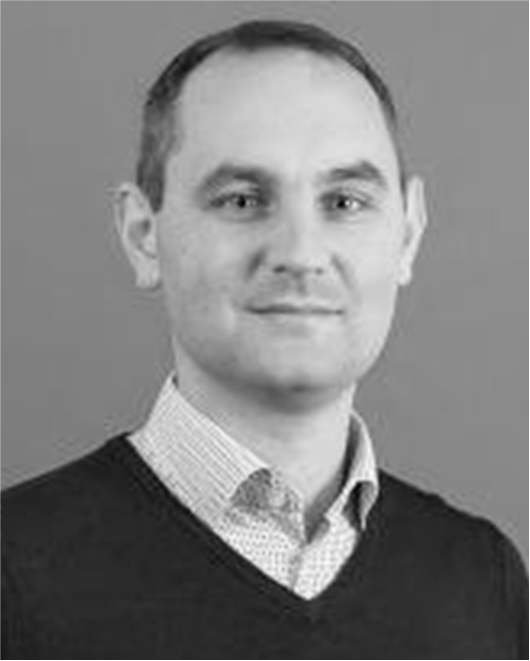}}]{Andrew Keane}(Senior Member, IEEE)
received the Ph.D. degree in electrical engineering from University College Dublin (UCD), Dublin, Ireland, in 2007. He is a Professor and Director of the Energy Institute with UCD. His research interests include power systems planning and operation, distributed energy resources, and distribution networks.
\end{IEEEbiography}

\vfill


\end{document}